\documentclass[proof]{pasj01}
\Received{$\langle$reception date$\rangle$}
\Accepted{$\langle$acception date$\rangle$}
\Published{$\langle$publication date$\rangle$}

\begin{document}

\title{{ 
Large-scale CO $J$=1--0 observations of the giant molecular cloud associated with the infrared ring N35 with the Nobeyama 45-m telescope
}}
\author{Kazufumi Torii\altaffilmark{1}, Shinji Fujita\altaffilmark{2}, Mitsuhiro Matsuo\altaffilmark{1}, Atsushi Nishimura\altaffilmark{2}, Mikito Kohno\altaffilmark{2}, Mika Kuriki\altaffilmark{3}, Yuya Tsuda\altaffilmark{5}, Tetsuhiro Minamidani\altaffilmark{1,6}, Tomofumi Umemoto\altaffilmark{1,6}, Nario Kuno\altaffilmark{3, 4}, Yusuke Hattori\altaffilmark{2}, Satoshi Yoshiike\altaffilmark{2}, Akio Ohama\altaffilmark{2}, Kengo Tachihara\altaffilmark{2}, Kazuhiro Shima\altaffilmark{7}, Asao Habe\altaffilmark{7}, Yasuo Fukui\altaffilmark{2}}%
\altaffiltext{1}{Nobeyama Radio Observatory, 462-2 Nobeyama Minamimaki-mura, Minamisaku-gun, Nagano 384-1305, Japan}
\altaffiltext{2}{Graduate School of Science, Nagoya University, Chikusa-ku, Nagoya, Aichi 464-8601, Japan}
\altaffiltext{3}{Department of Physics, Graduate School of Pure and Applied Sciences, University of Tsukuba, 1-1-1 Ten-nodai, Tsukuba, Ibaraki 305-8577, Japan}
\altaffiltext{4}{Tomonaga Center for the History of the Universe, University of Tsukuba, Tsukuba, Ibaraki 305-8571, Japan}
\altaffiltext{5}{Meisei University, 2-1-1 Hodokubo, Hino, Tokyo 191-0042, Japan}
\altaffiltext{6}{Department of Astronomical Science, School of Physical Science, SOKENDAI (The Graduate University for Advanced Studies), 2-21-1, Osawa, Mitaka, Tokyo 181-8588, Japan}
\altaffiltext{7}{Department of Physics, Faculty of Science, Hokkaido University, Kita 10 Nishi 8 Kita-ku, Sapporo 060-0810, Japan}

\email{kazufumi.torii@nao.ac.jp}

\KeyWords{ISM: clouds --- ISM: molecules --- radio lines: ISM --- stars: formation}

\maketitle

\begin{abstract}

We report an observational study of the giant molecular cloud (GMC) associated with the Galactic infrared ring-like structure N35 and two nearby H{\sc ii} regions G024.392+00.072 (H{\sc ii} region A) and G024.510-00.060 (H{\sc ii} region B), using the new CO $J$=1--0 data obtained as a part of the FOREST Unbiased Galactic Plane Imaging survey with the Nobeyama 45-m telescope (FUGIN) project at a spatial resolution of 21$''$. 
Our CO data revealed that the GMC, with a total molecular mass of $2.1\times10^6$\,$M_\odot$, has two velocity components over $\sim$10--15\,km\,s$^{-1}$. 
The majority of molecular gas in the GMC is included in the lower-velocity component (LVC) at $\sim$110--114\,km\,s$^{-1}$, while the higher-velocity components (HVCs) at $\sim$118--126\,km\,s$^{-1}$ consist of three smaller molecular clouds which are located near the three H{\sc ii} regions.  
The LVC and HVCs show spatially complementary distributions along the line-of-sight, despite large velocity separations of $\sim$5--15\,km\,s$^{-1}$, and are connected in velocity by the CO emission with intermediate intensities.
By comparing the observations with simulations, we discuss a scenario where collisions of the three HVCs with LVC at velocities of $\sim$10--15\,km\,s$^{-1}$ can provide an interpretation of these two observational signatures.
The intermediate velocity features between the LVC and HVCs can be understood as broad bridge features, which indicate the turbulent motion of the gas at the collision interfaces, while the spatially complementary distributions represent the cavities created in the LVC by the HVCs through the collisions. 
Our model indicates that the three H{\sc ii} regions were formed after the onset of the collisions, and it is therefore suggested that the high-mass star formation in the GMC was triggered by the collisions.
\end{abstract}

\section{Introduction}
It is of fundamental importance to understand the formation mechanism of high-mass stars in our long-term efforts to elucidate the structure and evolution of galaxies, as high-mass stars are influential in the interstellar medium (ISM); they inject a large amount of energy via ultraviolet (UV) radiation, stellar winds, and supernova explosions.

It is increasingly evident that cloud- cloud collision (CCC) plays an important role on formation of high-mass stars.
Observational studies of CCC were carried out in the high-mass star forming regions in the Milky Way (MW) and the Large Magellanic cloud (LMC) (e.g., \cite{lor1976, fur2009, tor2011, fuk2014, fuk2017b}), which include H{\sc ii} regions excited by a single O star (e.g., M20: \cite{tor2011,tor2017}, RCW120: \cite{tor2015}), Galactic super star clusters that have a few tens O stars within a small volume (e.g., Westerlund\,2: \cite{fur2009,oha2010}, NGC3603: \cite{fuk2014}, RCW38: \cite{fuk2016}), large H{\sc ii} regions extended for several tens of pc (e.g., W51: \cite{oku2001} and \cite{fuj2017ccc1} and NGC6334+NGC6357: \cite{fuk2017ccc2}), the GMCs in the LMC (e.g., \cite{fuk2015b,sai2017}), and so on.

In addition, theoretical studies of CCC have been conducted by many researchers (e.g., \cite{sto1970a, hab1992, ino2013, tak2014, haw2015b}).
\citet{hab1992} calculated a collision between two clouds with different sizes, followed by \citet{ana2010} and \citet{tak2014}, indicating that CCC can induce formation of dense self-gravitating clumps within a dense gas layer compressed by collision.
Formation of the massive clumps in the collisional-compressed layer was also discussed in depth in the magneto-hydrodynamical (MHD) simulations by \citet{ino2013} and \citet{ino2017}.
\citet{wu2017b} discussed that collision between GMCs increases star formation rate and efficiency.
Using a radiative transfer code, \citet{haw2015b} and \citet{haw2015} post-processed the CCC model data calculated by \citet{tak2014}, demonstrating observational signatures characteristic to CCC. 
Their discussion was confirmed by the molecular line observations by \citet{tor2017ccc}, \citet{tor2017}, and \citet{fuk2017ccc1} in the Galactic H{\sc ii} regions.
Global scale numerical calculations indicate that CCCs can be as frequent as every $\sim10$\,Myrs in a MW-like galaxy \citep{tas2009,dob2015}.

These studies have shown that CCC is a promising mechanism of the formation of massive dense clumps which lead to the formation of high-mass stars, and it is therefore of crucial importance to enlarge the samples of the CCC regions for comprehensive understanding of CCC and subsequent high-mass star formation in the MW. 
For such a purpose, infrared ring or bubbling structures distributed in the Galactic plane are important observational targets. Based on the Spitzer infrared observations, \citet{chu2006} and \citet{chu2007} identified about 600 ring-like 8\,$\micron$ structures in the Galactic plane; this was followed by an expanded catalog of $\sim$5100 rings by \citet{sim2012}. 
\citet{deh2010} and \citet{ken2012} discussed that the majority of the identified infrared ring-like structures are associated with the H{\sc ii} regions embedded within the molecular clouds.
These infrared ring-like structures therefore provide important sites in which to study high-mass star formation, and molecular line observations in the several infrared ring-like structures,  i.e., RCW\,120 (N9 in \cite{chu2006}) by \cite{tor2015}, N37 by \citet{bau2016}, N18 by \citet{tor2017ccc}, N49 by \citet{dew2017b}, and S116--118 by \citet{fuk2017ccc3}, have indicated that these high-mass star forming regions were likely formed by the triggering of CCC.

\begin{figure}
 \begin{center}
  \includegraphics[width=10cm]{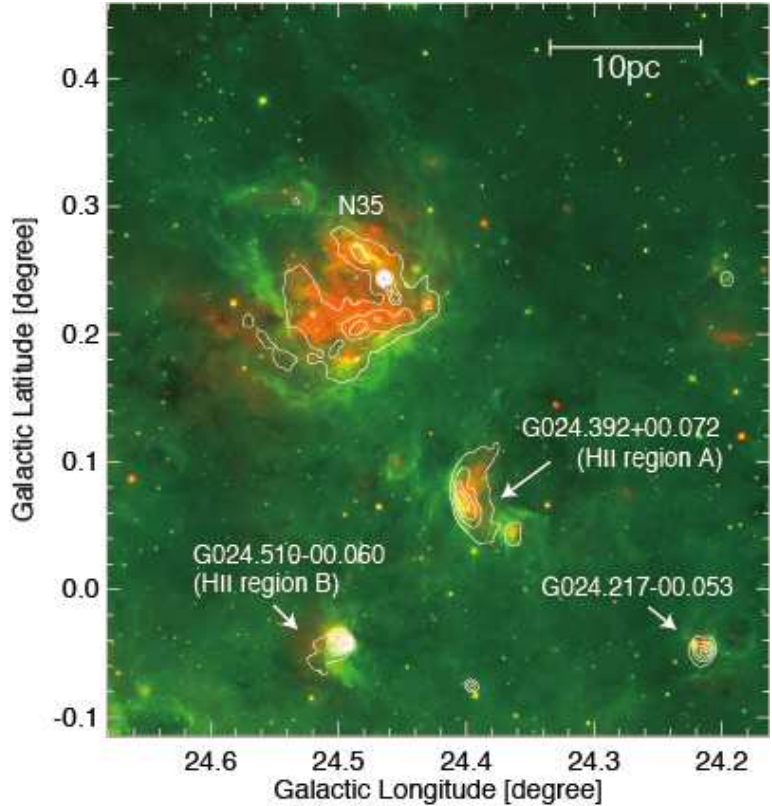}
 \end{center}
 \caption{A composite color image of N35 and nearby H{\sc ii} regions, where green shows the {\it Spitzer}/GLIMPSE 8\,$\micron$ emission \citep{ben2003, chu2009}, while red is the {\it Spitzer}/MIPSGAL 24\,$\micron$ emission \citep{car2009}. Contours indicate the THOR 21\,cm radio continuum data \citep{beu2016}. The scale-bar is plotted assuming a distance of 8.8\,kpc.}\label{fig:rgb}
\end{figure}

In this paper, we report an observational study of the GMC which includes the infrared ring-like structure N35 and nearby H{\sc ii} regions with the CO $J$=1--0 data obtained using the Nobeyama 45-m telescope.
Figure\,\ref{fig:rgb} shows a two color composite image of the {\it Spitzer}/GLIMPSE 8\,$\mu$m emission (green, \cite{ben2003, chu2009}) and the {\it Spitzer}/MIPSGAL 24\,$\mu$m emission (red, \cite{car2009}) for a large area including N35 and nearby H{\sc ii} regions, where the contours indicate the H{\sc i}/OH/Recombination line survey of the Milky Way (THOR) 21\,cm radio continuum data, which has a beam size of $\sim25''$ \citep{beu2016}. 
The 24\,$\micron$ emission, which is attributed to warm dust grains, roughly coincides with the 21\,cm distribution, while the 8\,$\micron$ emission dominated by polycyclic aromatic hydrocarbons (PAHs) surrounds the 21\,cm and 24\,$\micron$ distributions. 
Observations of the radio recombination line indicates a radial velocity of the ionized gas in N35 as $\sim$115.7$\pm$1.2\,km\,s$^{-1}$ \citep{loc1989}.

Other than N35, there are three H{\sc ii} regions at the south to N35 as shown in Figure\,\ref{fig:rgb}, i.e., G024.392+00.072, G024.510-00.060, and G024.217-00.053 \citep{and2014}. 
Radio recombination line observations indicate that the former two have radial velocities of $\sim$110.0$\pm$1.2 \citep{loc1989} and $\sim$108.1$\pm$0.5\,km\,s$^{-1}$ \citep{sew2004}, respectively, which are similar to that in N35, while the last one has a velocity of $\sim$82.0$\pm$2.4\,km\,s$^{-1}$ \citep{loc1989}, suggesting that it is located at a different distance from N35. 
For the sake of convenience, we hereafter refer the former two H{\sc ii} regions, G024.392+00.072 and G024.510-00.060, as ``H{\sc ii} region A'' and ``H{\sc ii} region B'', respectively. 

The observed radial velocities of N35 and H{\sc ii} regions A and B, $\sim$108--115\, km\,s$^{-1}$, correspond to the near- and far-kinematic distances of 6.4--6.9\,kpc and 8.6--9.1\,kpc, respectively \citep{and2009b}. 
Based on the absorption and self-absorption studies of the H{\sc i} 21\,cm profiles, \citet{and2009b} favored the far-kinematic distances in these three H{\sc ii} regions.
Following their results, in this paper we adopt the common distance of 8.8\,kpc to the three H{\sc ii} regions with an uncertainty of 0.3\,kpc, as all of these H{\sc ii} regions are associated with a single GMC as presented later in Section 3 with our new CO data. 
Assuming a distance of 8.8\,kpc, the size of the H{\sc ii} region in N35 is measured as $7'\simeq18$\,pc, and the total infrared luminosity of N35 measured by \citet{hat2016} can be rescaled to $\sim10^{6.7}\,L_\odot$.

Molecular line observations of N35 was performed by \citet{bea2010} in the $^{12}$CO $J$=3--2 transition using the James Clerk Maxwell Telescope (JCMT) telescope for a $\sim15'\times15'$ area centered on N35, indicating that molecular gas at $\sim$110--120\,km\,s$^{-1}$, although detailed distribution and dynamics of molecular gas for a large area of N35 has not been studied.
In this study, we present the molecular gas distribution for a $0\fdg52\times0\fdg57$ region including N35 and H{\sc ii} regions A and B using the CO $J$=1--0 data obtained with the Nobeyama 45-m radio telescope at an angular resolution of 21$''$, which corresponds to $\sim$0.9\,pc at 8.8\,kpc. 
The high spatial resolution provides a wealth of information on distribution and dynamics of the molecular gas, allowing us to investigate high-mass star formation in this region. In section\,2 we describe the CO $J$=1--0 dataset used in this study, and in section\,3 we present the main results of the analyses of the CO dataset and comparisons with the other wavelengths. In section\,4 we discuss the results and a summary is presented in section\,5. In this paper, we refer to the four points of the compass based on the Galactic coordinates.

\section{Data set}
The $^{12}$CO, $^{13}$CO, and C$^{18}$O $J$=1--0 dataset presented in this paper was obtained as a part of FOREST Unbiased Galactic plane Imaging survey with the Nobeyama 45-m telescope (FUGIN; for the full description of the observations and data reduction, see \cite{ume2017}). 
FUGIN is a large scale Galactic plane survey of the three CO isotopologues using the FOur-beam REceiver System on the 45-m Telescope receiver (FOREST; \cite{min2016}), which is a four beams, dual-polarizaiton, and two sideband receiver. 
Typical system temperatures of FOREST were $\sim$250\,K for $^{12}$CO and $\sim$150\,K for $^{13}$CO and C$^{18}$O.
The backend system was the digital spectrometer ``SAM45'', which provided a bandwidth of 1\,GHz and a resolution of 244.14\,kHz, which correspond to $\sim$2600\,km\,s$^{-1}$ and $\sim$1.3\,km\,s$^{-1}$, respectively, at 115\,GHz.  
The observations were made in the on-the-fly mode with an output grid size of 8.5$''$ in space and 0.65\,km\,s$^{-1}$ in velocity. 
We smoothed the output data with a two-dimensional Gaussian function to a spatial resolution of 25$''$ to improve sensitivities. 
The final root-mean-square (r.m.s) noise fluctuations of the data were 0.8\,K for $^{12}$CO and 0.4\,K for $^{13}$CO and C$^{18}$O at a channel resolution of 0.65\,km\,s$^{-1}$.

\section{Results}
\subsection{Large-scale CO distributions}
\begin{figure}
 \begin{center}
  \includegraphics[width=16cm]{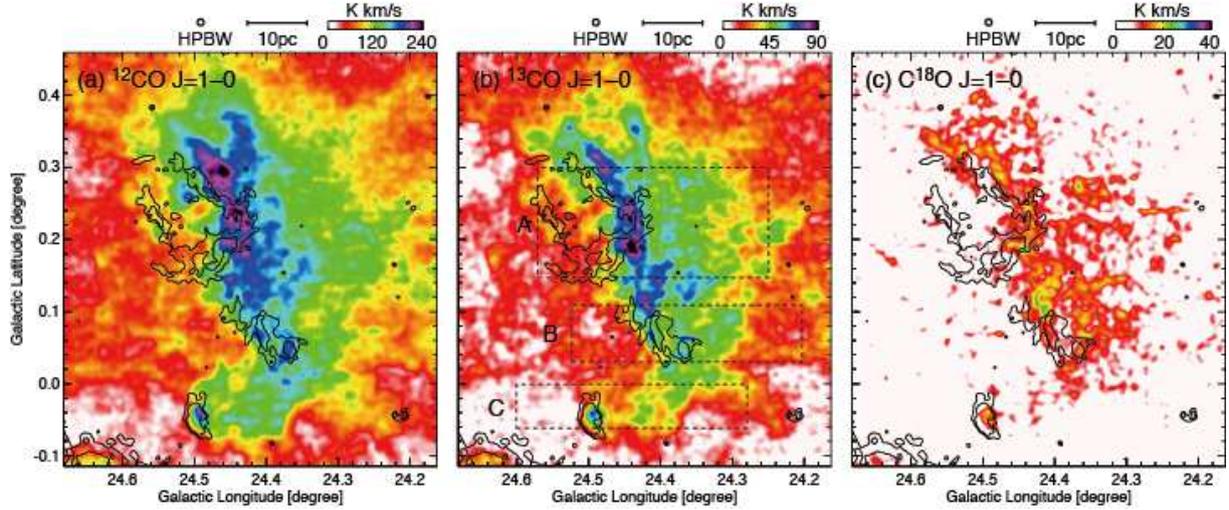}
 \end{center}
 \caption{(a) Integrated intensity distributions of the FUGIN CO data for a velocity range of 105--125\,km\,s$^{-1}$. The maps of $^{12}$CO, $^{13}$CO, and C$^{18}$O $J$=1--0 are presented in (a), (b), and (c), respectively. Contours indicate the 8\,$\micron$ emission plotted at 150 and 200\,MJy\,str$^{-1}$. Dashed lines in (b) indicate the areas used for the position-velocity diagrams shown in Figure\,\ref{fig:lv}. }\label{fig:wco}
\end{figure}

Figure\,\ref{fig:wco} shows intensity distributions of the three CO lines integrated over a velocity range of $\sim$105--125\,km\,s$^{-1}$, at which the CO emission associated with N35 and H{\sc ii} regions A and B is pronounced. 
We identified a GMC in this velocity range which has a size of about 30\,pc\,$\times$\,50\,pc at 8.8\,kpc. We hereafter refer the GMC as the ``N35 GMC''.
As shown in Figures\,\ref{fig:wco}(a) and (b), the $^{12}$CO and $^{13}$CO emission in the N35 GMC are enhanced at the eastern rim of the cloud at $l\sim24\fdg4$--$24\fdg5$ and $b\sim0\fdg05$--$0\fdg35$, forming a ridge feature which is stretched along the north-south direction.
N35 is located on the east of this ridge feature, while H{\sc ii} region A is distributed around the southern end of the ridge feature.
H{\sc ii} region B is seen at the southeastern rim the GMC.
C$^{18}$O is widely detected in the N35 GMC (Figure\,\ref{fig:wco}(c)), indicating the presence of dense gas.

By assuming an X(CO)-factor of $2\times10^{20}$\,cm$^{-2}$\,(K\,km\,s$^{-1}$)$^{-1}$ \citep{str1998}, which is the CO-to-H$_2$ conversion factor, the total molecular mass of the GMC can be calculated as $\sim2.1\times10^6\,M_\odot$ from the $^{12}$CO map in Figure\,\ref{fig:wco}(a), where we defined the GMC by drawing a
contour at 70\,K\,km\,s$^{-1}$. In addition, we measured the total molecular mass of the GMC from the $^{13}$CO $J$=1--0 data assuming local thermodynamic equilibrium. We estimated the excitation temperature $T_{\rm ex}$ of the $^{13}$CO emission at each
direction of the defined GMC with the following equation,
\begin{equation}
T_{\rm ex} \ {\rm (K)} \ = \ \frac{5.53}{\ln\{ 1 + 5.53 / (T(^{12}{\rm CO}) + 0.819) \} }, \label{eq0}
\end{equation}
where $T(^{12}{\rm CO})$ is the peak temperature of the $^{12}$CO profile. 
The derived $T_{\rm ex}$ ranges between 15 and 25\,K, and if we assume an abundance ratio between $^{12}$C and $^{13}$C of 55 \citep{wil1994}, the total molecular mass of the GMC can be estimated as $\sim1.4\times10^6$\,$M_\odot$, which is nearly consistent with the mass estimated from the $^{12}$CO data.
The H$_2$ column densities $N({\rm H}_2)$ tend to increase to (4--6)\,$\times10^{22}$\,cm$^{-2}$ at the regions that have $^{13}$CO integrated intensities larger than $\sim$70\,K\,km\,s$^{-1}$ in the $^{13}$CO map in Figure\,\ref{fig:wco}(b), which includes the ridge feature and the gas associated with H{\sc ii} region B, while the other part of the GMC that have lower $^{13}$CO integrated intensities show typical $N({\rm H}_2)$ of (2--3)\,$\times10^{22}$\,cm$^{-2}$.

\begin{figure}
 \begin{center}
  \includegraphics[width=16cm]{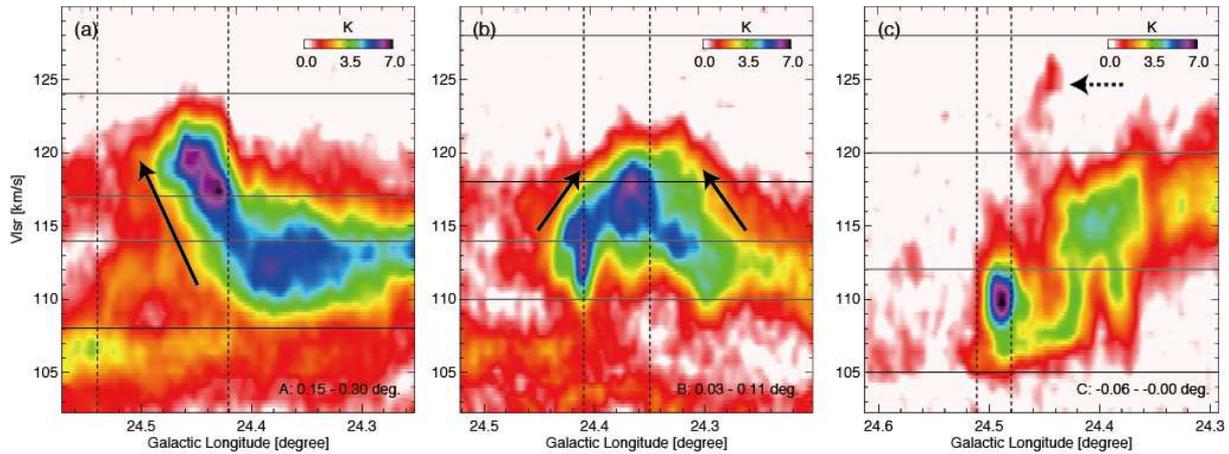}
 \end{center}
 \caption{Position-velocity diagrams of the $^{13}$CO emission toward the three H{\sc ii} regions, (a) N35, (b) H{\sc ii} region A, and (c) H{\sc ii} region B. Integration ranges in the Galactic latitude are plotted in Figure\,\ref{fig:wco}(b) with dashed lines. Horizontal solid lines indicate the velocity ranges shown in Figures\,\ref{fig:3vel1}--\ref{fig:3vel3}, while vertical dashed lines indicate the extents of the individual H{\sc ii} regions measured in the 21\,cm radio continuum map in Figure\,\ref{fig:rgb}. The arrows with solid lines in (a) and (b) indicate the velocity gradient of molecular gas, while the arrow with dashed lines in (c) depicts the HVC\,3. }\label{fig:lv}
\end{figure}

Figure\,\ref{fig:lv} shows the longitude-velocity diagrams of the $^{13}$CO emission around the three H{\sc ii} regions, N35 and H{\sc ii} regions A and B, where the extents of the H{\sc ii} regions measured with the 21\,cm map in Figure\,\ref{fig:rgb} are indicated by vertical dashed lines.
The integration ranges of these three diagrams are indicated as dashed lines in Figure\,\ref{fig:wco}(b) with labels A--C.
Figure\,\ref{fig:lv}(a) shows a steep velocity gradient of gas at the western part of N35 from $(l,v)\sim(24\fdg40, 110$\,km\,s$^{-1}$) to $(24\fdg45, 121$\,km\,s$^{-1}$), which corresponds $\sim1.4$\,km\,s$^{-1}$\,pc$^{-1}$.
Velocity gradients are also seen in the gas around H{\sc ii} region A. 
As indicated by the arrows, there are two velocity gradients. The CO emission is continuously distributed from $(l,v)\sim(24\fdg30, 112$\,km\,s$^{-1}$) to $(l,v)\sim(24\fdg42, 112$\,km\,s$^{-1}$) via $(l,v)\sim(24\fdg36, 118$\,km\,s$^{-1}$), forming an inverted V-shaped velocity distribution of gas. The velocity gradient seen in each side of the inverted V-shaped is measured as $\sim \pm0.7$\,km\,s$^{-1}$\,pc$^{-1}$, and H{\sc ii} region A is distributed at the eastern side of the inverted V-shaped at $l\sim24\fdg35$--$24\fdg41$. 
On the other hand, although we found no velocity gradient of gas around H{\sc ii} region B in Figure\,\ref{fig:lv}(c), we identified a spatially compact emission just west to H{\sc ii} region B at $(l,v)\sim(24\fdg45, 125$\,km\,s$^{-1}$) as indicated by the arrow with dashed line. 
The compact emission is connected to the molecular gas around 110\,km\,s$^{-1}$ by the diffuse CO emission at the intermediate velocities.
These position-velocity diagrams suggest that the N35 GMC has multiple velocity components, and these velocity components are connected with each other by the intermediate velocity emission.

\begin{figure}
 \begin{center}
  \includegraphics[width=16cm]{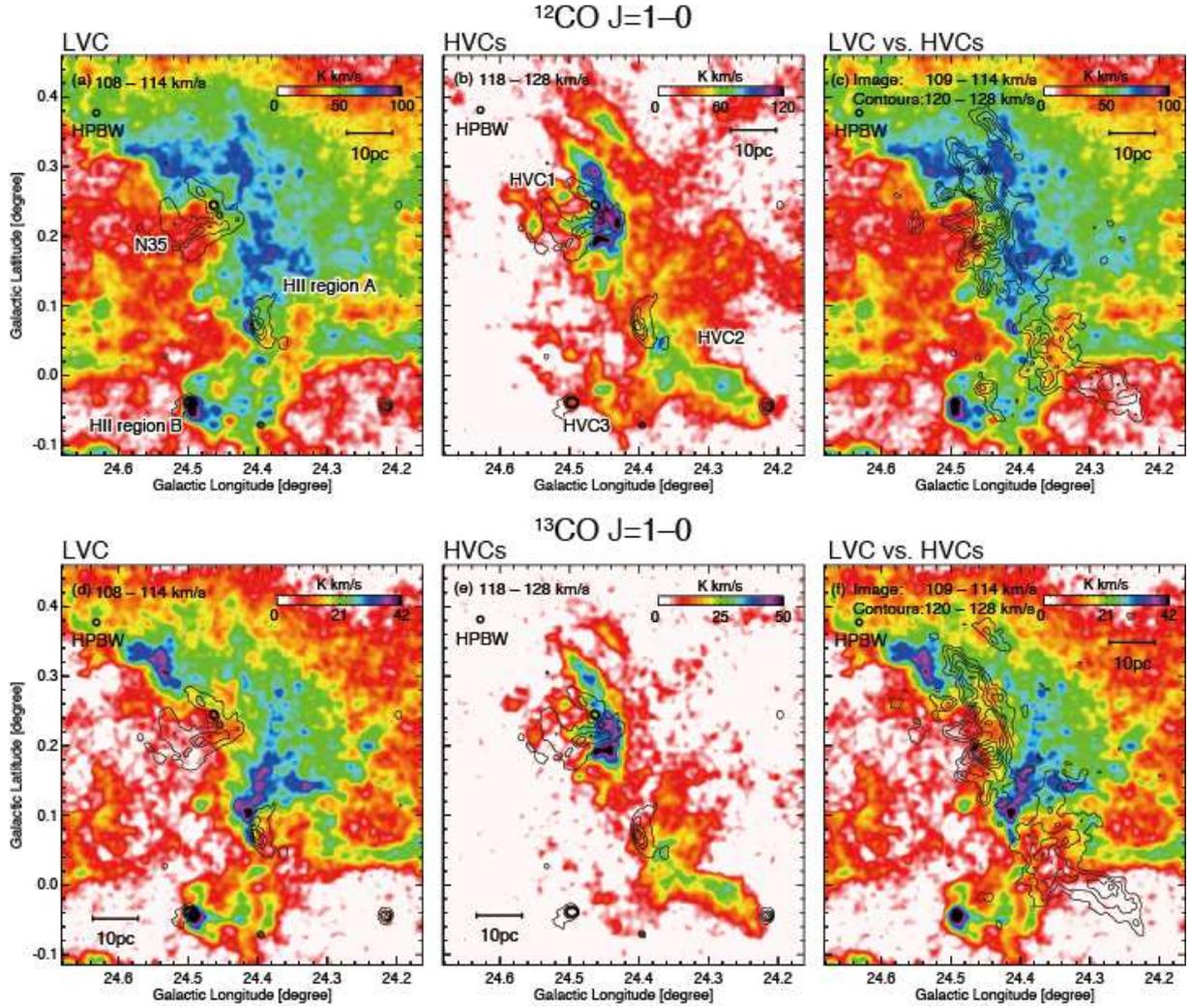}
 \end{center}
 \caption{Integrated intensity distributions of the LVC and HVCs in the N35 GMC. The upper and lower panels indicate $^{12}$CO and $^{13}$CO, respectively. The LVC is presented in the left panels (a) and (d) for $\sim$108--114\,km\,s$^{-1}$, while the HVCs are in the center panels (b) and (e) for $\sim$118--128\,km\,s$^{-1}$. The contours in the left and center panels indicate the 21\,$\micron$ radio continuum map, starting at 3\,$\sigma$ with steps of 1.5$\sigma$, which respectively corresponds to 45 and 22.5\,mJy\,beam$^{-1}$. The right panels (c) and (f) show the comparisons between the LVC (image) and HVCs (contours), where the contours are plotted every 10\,K\,km\,s$^{-1}$ from 21\,K\,km\,s$^{-1}$ in $^{12}$CO in (c) and every 5\,K\,km\,s$^{-1}$ from 7\,K\,km\,s$^{-1}$ in $^{13}$CO in (f).}\label{fig:2vel}
\end{figure}

The spatial distributions of the two velocity components are presented in Figure\,\ref{fig:2vel} for two velocity ranges, 108--114\,km\,s$^{-1}$ and 118--128\,km\,s$^{-1}$, where the upper and lower panels of the figure show the $^{12}$CO and $^{13}$CO emission, respectively.
We also present the velocity channel maps of the $^{12}$CO, $^{13}$CO, and C$^{18}$O emissions for a velocity range from $\sim102$\,km\,s$^{-1}$ to $\sim125$\,km\,s$^{-1}$ in Figures\,\ref{fig:12channel}--\ref{fig:18channel} in the Appendix as a supplement.
In Figures\,\ref{fig:2vel}(a) and (d) it is seen that the majority of molecular gas in the N35 GMC is included in the lower velocity range, while in the higher velocity range in Figures\,\ref{fig:2vel}(b) and (e) there are mainly two molecular gas components toward N35 and the southwest of H{\sc ii} region B, which respectively correspond to the higher velocity parts of the velocity gradients seen in the position-velocity diagrams in Figures\,\ref{fig:lv}(a) and (b).
In addition, the compact CO emission seen at the west of H{\sc ii} region B in Figure\,\ref{fig:lv}(c) is seen in the higher velocity range shown in Figures\,\ref{fig:2vel}(b) and (e).
We hereafter refer the larger molecular gas component in the lower velocity range as the ``LVC (lower velocity component)'', while the gas components in the higher velocity components the ``HVCs (higher velocity components)''.
The three HVCs distributed in N35, H{\sc ii} region A, and H{\sc ii} region B are referred as the ``HVC\,1'', ``HVC\,2'', and ``HVC\,3'', respectively, as dubbed in Figure\,\ref{fig:2vel}(b).

In Figures\,\ref{fig:2vel}(a) and (d)  it is seen that the majority of molecular gas in the N35 GMC is included in the lower velocity range, while in the higher velocity range in Figures\,\ref{fig:2vel}(b) and (e)  there are mainly two molecular gas components toward N35 and the southwest of H{\sc ii} region B, which respectively correspond to the higher velocity parts of the velocity gradients seen in the position-velocity diagrams in Figures\,\ref{fig:lv}(a) and (b). 
In addition, the compact CO emission seen at the west of H{\sc ii} region B in Figure\,\ref{fig:lv}(c) is seen in the higher velocity
range shown in Figures\,\ref{fig:2vel}(b) and (e). 
We hereafter refer to the larger molecular gas component in the lower velocity range as the ``LVC (lower-velocity component)'', and to the gas components in the higher-velocity range as the ``HVCs (higher velocity components)''. 
The three HVCs distributed in N35, H{\sc ii} region A, and H{\sc ii} region B are referred to as ``HVC\,1'', ``HVC\,2'', and ``HVC\,3'', respectively, as dubbed in Figure\,\ref{fig:2vel}(b).

In N35 the CO emission in the LVC appears to surround the 21\,cm radio continuum map (Figures\,\ref{fig:2vel}(a) and (d)), whereas the HVC\,1 which is elongated along the north-south direction coincides with the 21\,cm distribution (Figures\,\ref{fig:2vel}(b) and (e)).
The LVC and HVC1 exhibit complementary distribution, as shown in figures Figures\,\ref{fig:2vel}(c) and (f). 
The steep velocity gradient seen in the position-velocity diagram in Figure\,\ref{fig:lv}(a) is distributed at the interface of the complementary distribution. 
Complementary distribution of gas between the LVC and HVCs is also seen around H{\sc ii} regions A and B. 
The LVC shows an intensity depression toward H{\sc ii} region A, whereas the HVC\,2 overlaps the 21\,cm emission
of H{\sc ii} region A with a shape elongated toward the southwest, showing complementary distribution between these two velocity components. Similar to the case in N35, the velocity gradients seen Figure\,\ref{fig:lv}(b) are distributed around
the interface of the complementary distribution. 
The HVC\,3 shows a compact circular distribution with radius $\sim$3\,pc at $(l,b)\sim(24\fdg45, -0\fdg02)$, and is spatially coincident with the central hole of a ring-like molecular structure in the LVC,
the outer radius of which is measured as $\sim$7--8\,pc, indicating complementary distribution between the LVC and HVC3.
H{\sc ii} region B is located in the southeast of the ring-like structure of the LVC.

\begin{figure}
 \begin{center}
  \includegraphics[width=12cm]{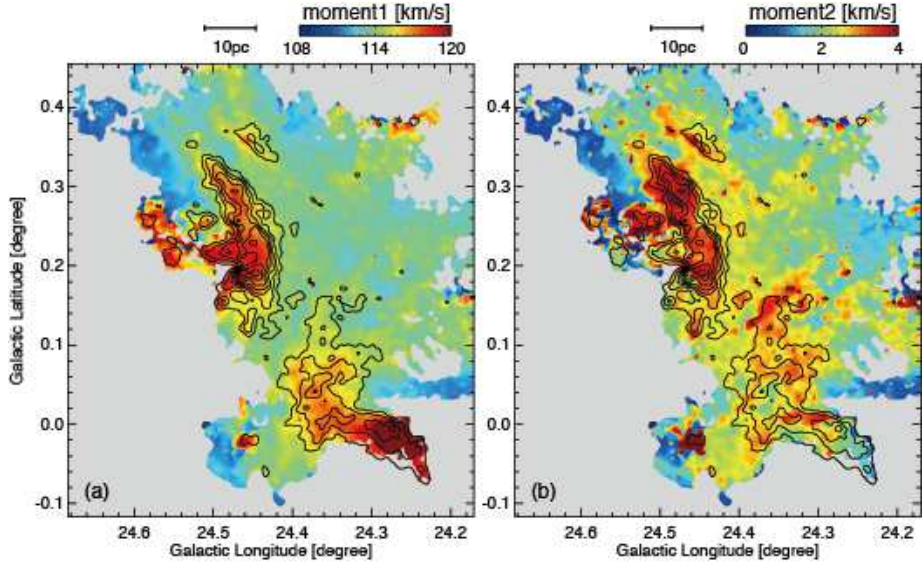}
 \end{center}
 \caption{The first and second moment maps the $^{13}$CO data are presented in (a) and (b), respectively. These two maps were both created for a velocity range of 105--125\,km\,s$^{-1}$ using the volume pixels (voxels) with $^{13}$CO intensities of higher than 1\,K, where the data were smoothed to a velocity resolution of 1.95\,km\,s$^{-1}$ to reduce noise. The contours indicate the $^{13}$CO integrated intensity distributions of the HVCs in 120--128\,km\,s$^{-1}$, and are plotted at the same levels as those in Figure\,\ref{fig:2vel}(f).}\label{fig:mom1}
\end{figure}

In Figure\,\ref{fig:mom1} we plot the first and second moment maps of the $^{13}$CO data, where the $^{13}$CO contour maps of the HVCs are superimposed. 
In the first moment map in Figure\,\ref{fig:mom1}(a) it is seen that the three HVCs have velocities of $\sim$117--120\,km\,s$^{-1}$. 
While the HVC\,1 shows nearly uniform velocities around 118\,km\,s$^{-1}$, the HVC\,2 shows a velocity gradient along its elongated shape, with the velocity peaked around its southwestern end. 
In the second moment map in Figure\,\ref{fig:mom1}(b), on the other hand, CO emission that have broad velocity widths larger than 3\,km\,s$^{-1}$ is detected around the HVCs. These broad velocity features indicate the intermediate-velocity gas components connecting the LVC and HVCs in velocity, which include the steep velocity gradients seen in the position-velocity diagrams of the HVC\,1 and HVC\,2 in Figures\,\ref{fig:lv}(a) and (b), respectively.
The obtained velocity dispersions of these intermediate velocity features are significantly larger than the thermal velocity dispersion of the molecular gas, $\sim$1\,km\,s$^{-1}$, even if we assume a high gas temperature of 30\,K from the molecular line observations of the Galactic H{\sc ii} regions (e.g., \cite{tor2011, fuk2016}), suggesting enhancement of turbulent motion of gas in these regions.

\subsection{Detailed gas distributions in the three H{\sc ii} regions}
In this subsection we present detailed CO distributions of each of the three H{\sc ii} regions, N35, H{\sc ii} region A, and H{\sc ii} region B, including comparisons with the infrared images.

\subsubsection{N35}
\begin{figure}
 \begin{center}
  \includegraphics[width=13cm]{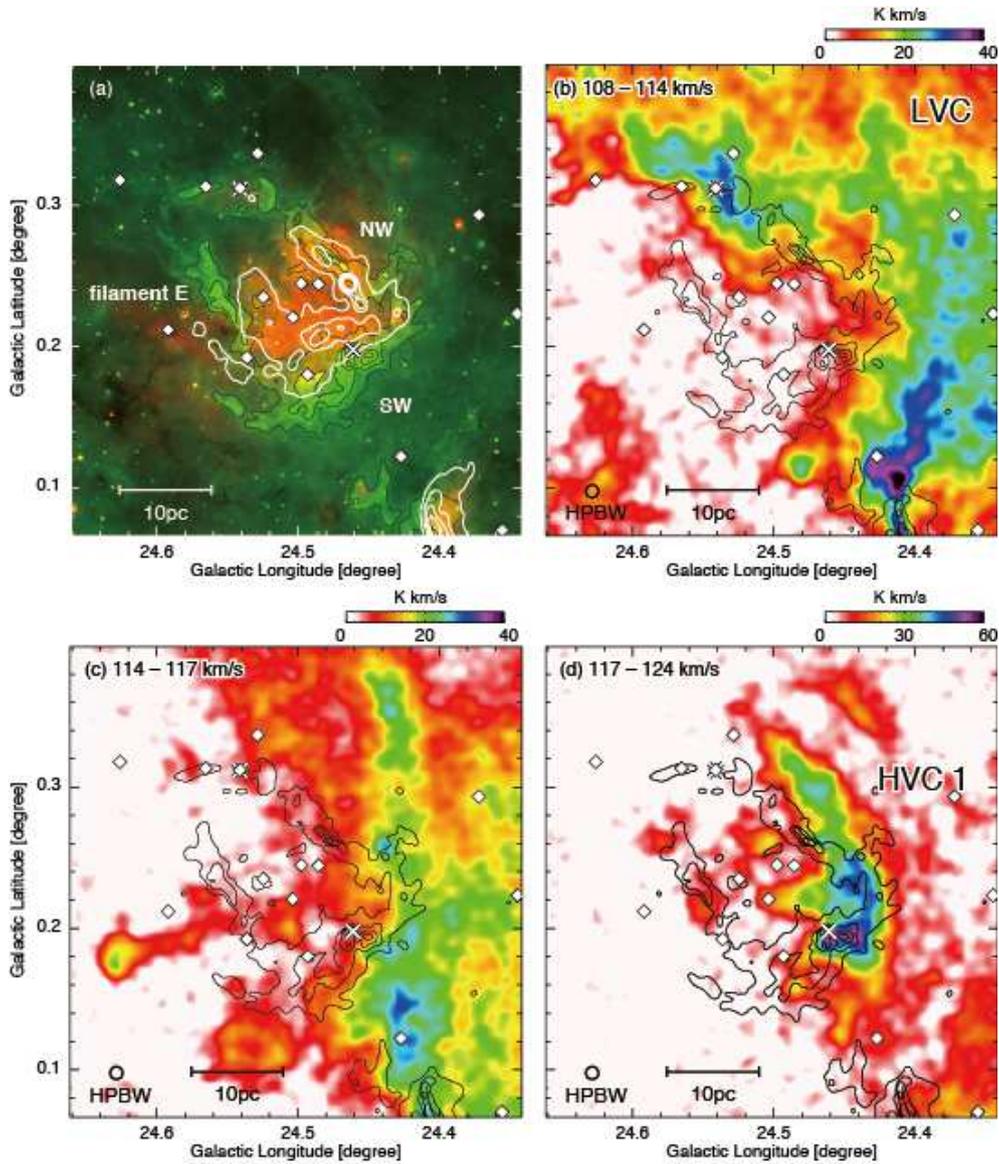}
 \end{center}
 \caption{(a) A two color composite image of N35. Green and red show the {\it Spitzer} 8\,$\micron$ and 24\,$\micron$ data, respectively. The black green contours indicate the 8\,$\micron$ emission, which are plotted at 150, 200, 250, and 300\,MJy\,str$^{-1}$, while the white contours show the THOR 21\,cm radio continuum emission plotted at the same levels as those in Figure\,\ref{fig:rgb}. (b--d) Integrated intensity distributions of the $^{13}$CO emission in N35 for three velocity ranges, (b) 110--114\,km\,s$^{-1}$, (c) 114--118\,km\,s$^{-1}$, and (d) 118--124\,km\,s$^{-1}$. These velocity ranges are plotted with solid lines in Figure\,\ref{fig:lv}(a). The black contours indicate the 8\,$\micron$ emission plotted at the same level as (a). The crosses depict the CH$_3$OH masers identified by \citet{bre2015}, while the diamonds show the {\it Spitzer} red sources cataloged by \citet{rob2008}.}\label{fig:3vel1}
\end{figure}

Figure\,\ref{fig:3vel1}(a) shows an enlarged image of N35 in the {\it Spitzer} 8\,$\micron$ and 24\,$\micron$ emission.
The 8\,$\micron$ emission in N35 mainly consists of three filamentary structures in the east, northwest, and southwest. 
We hereafter refer to these three filaments as the filaments E, NW, and SW, respectively.
Although the exciting source of ionized gas in N35 has not been identified to-date, it is reasonable to assume that it is located around strong and compact peak of the 21\,cm emission at $(l,b)\sim(24\fdg46, 0\fdg25)$.
We obtained the 20 and 21\,cm radio flux densities $f_{\rm \nu}$ of N35 as $\sim$10.1\,Jy and $\sim$7.9\,Jy, respectively, from the web site of the Wide-field Infrared Survey Explorer (WISE) catalog of Galactic H{\sc ii} regions \footnote{http://astro.phys.wvu.edu/wise/}. 
The $f_{\rm \nu}$ at these two wavelengths were measured using the Multi-Array Galactic Plane Imaging Survey (MAGPIS) 20\,cm data \citep{hel2006} and the Very Large Array (VLA) Galactic Plane Survey (VGPS) 21\,cm data \citep{sti2006}, respectively (see \cite{mak2017} for details). 
The authors discussed that their estimates for $\sim$1000 sources have fractional errors of $\sim$13--49\% for the MAGPIS data and 30--92\% for the VGPS data.
Using the same MAGPIS data, \citet{bea2010} applied a Bayesian technique to derive the $f_{\rm \nu}$ of N35 as 7.57\,Jy, nearly the same as the estimate in the WISE catalog.

Based on the $f_{\rm \nu}$ in the WISE catalog, we computed the flux of Lyman continuum photons $N_{\rm Ly}$ with the following equation \citep{rub1968}:
\begin{equation}
N_{\rm Ly} \simeq 6.3\times10^{52} \left( \frac{T_{\rm e}}{10^4\,{\rm K}} \right)^{-0.45} \left( \frac{\nu}{{\rm GHz}} \right)^{0.1} \left( \frac{L_{\rm \nu}}{10^{20}\,{\rm WHz^{-1}}} \right), \label{eq:1} \\
\end{equation}
where $L_{\rm \nu} \equiv 4\pi D^2 f_\nu$ and $D$ is the distance to N35, 8.8\,kpc. 
We assumed an electron temperature $T_{\rm e}$ of 8000\,K.
The $N_{\rm Ly}$ of N35 were derived as $\sim10^{49.83}$\,s$^{-1}$ for the MAGPIS 20\,cm data and $\sim10^{49.72}$\,s$^{-1}$ for the VGPS 21\,cm data as summarized in Table\,\ref{tab:1}. 
These figures indicate that the exciting source of N35 has a spectral type of earlier than O3V \citep{mar2005}, if we assume a single object.
Note that if we vary the distance between 8.6 and 9.1\,kpc, the derived $N_{\rm Ly}$ is changed by $\pm\sim5\%$.

Detailed CO distributions in N35 are shown in Figures\,\ref{fig:3vel1}(b), (c), and (d) for three velocity ranges, 108--114, 114--117, and 117--124\,km\,s$^{-1}$, respectively.
The first and last velocity ranges, in Figures\,\ref{fig:3vel1}(a) and (c), cover the LVC and HVC\,1, respectively, as indicated by horizontal lines in Figure\,\ref{fig:lv}(a), while the intermediate-velocity gas is shown in the second velocity range in Figures\,\ref{fig:3vel1}(b). 
In Figure\,\ref{fig:3vel1} we plotted the {\it Spitzer} red sources identified by \citet{rob2008} with diamonds, as well as the CH$_3$OH maser sources with crosses \citep{bre2015}. 
Note that a statistical study by \citet{rob2008} indicated that the cataloged {\it Spitzer} red sources includes 50--70\% of YSOs and 30--50\% of asymptotic giant branch stars. 

In Figure\,\ref{fig:3vel1}(b) the LVC appears to surround the filament NW and SW, while the HVC\,1 in Figure\,\ref{fig:3vel1}(d) has a vertically elongated bow-like structure, which harbors a clumpy structure at the corresponding direction of the bright 8\,$\micron$ spot within the filament SW at $(l,b)\sim(24\fdg46, 0\fdg20)$.
A CH$_3$OH maser source is located just at the east of the peak, implying massive star formation.
There is another CH$_3$OH maser source around the northern end of the filament NW at $(l,b)\sim(24\fdg54, 0\fdg31)$, which is spatially coincident with a {\it Spitzer} red source. 
The molecular counterpart of these sources are detected in the LVC in Figure\,\ref{fig:3vel1}(b)
These observed signatures indicate that the LVC and HVC\,1 are physically associated with N35, and the complementary distribution and the steep velocity gradient between the LVC and HVC\,1 suggest physical interaction between these two.
In the intermediate-velocity range in Figure\,\ref{fig:3vel1}(c), the CO emission shows a vertically straight feature at $l\sim24\fdg43$. 
The northern half of the straight feature corresponds to the diffuse 8\,$\micron$ emission seen at $b>0\fdg25$ (Figure\,\ref{fig:3vel1}(a)), while the southern part shows a possible association with a {\it Spitzer} red source at $b \sim 0\fdg2$.

\begin{figure}
 \begin{center}
  \includegraphics[width=13cm]{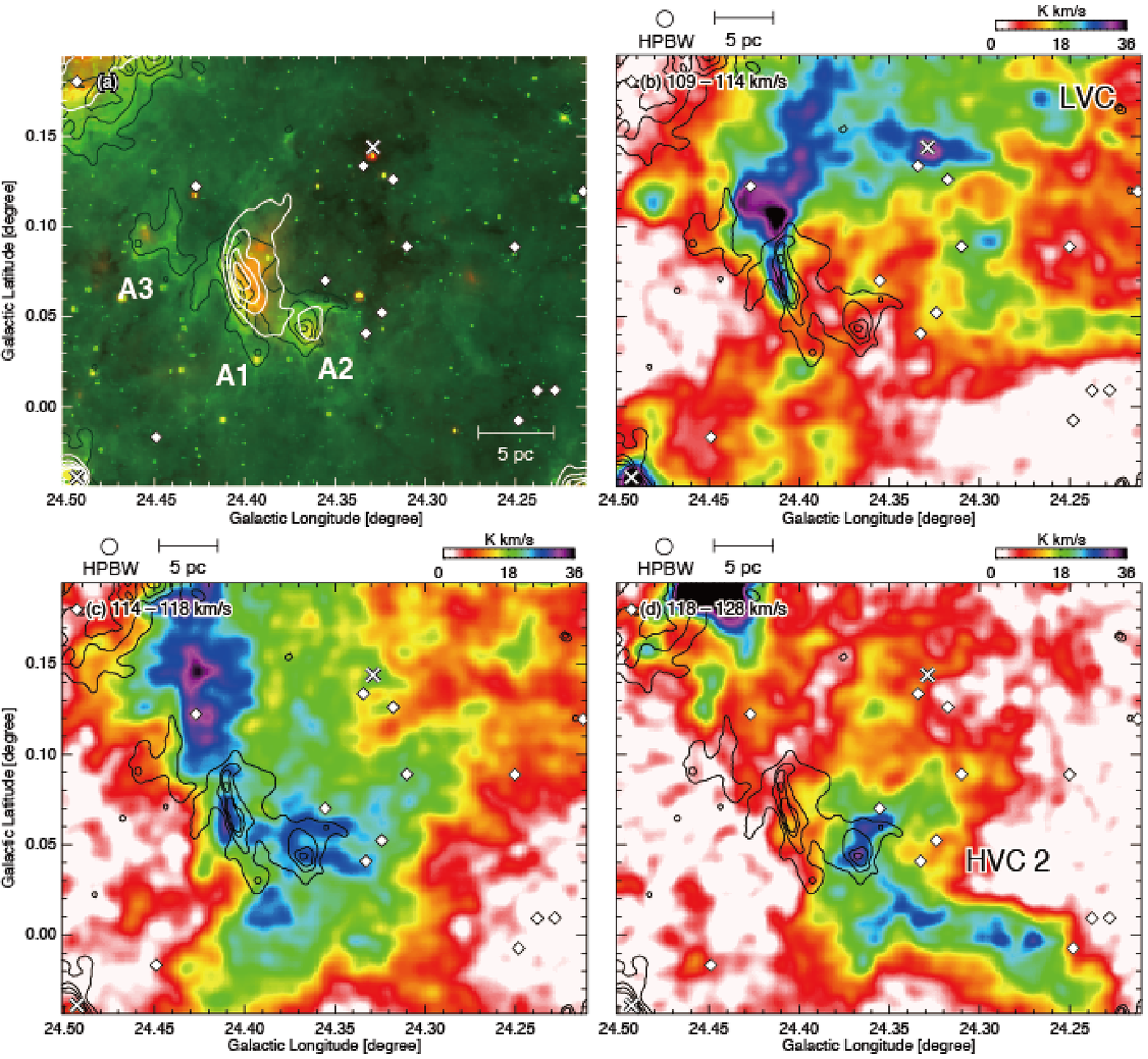}
 \end{center}
 \caption{Same as Figure\,\ref{fig:3vel1} but for H{\sc ii} region A. The velocity ranges in (b), (c), and (d) are 109--114\,km\,s$^{-1}$, 114--120\,km\,s$^{-1}$, and 120--128\,km\,s$^{-1}$, respectively.}\label{fig:3vel2}
\end{figure}

\subsubsection{H{\sc ii} region A}
Detailed CO distributions around H{\sc ii} region A are shown in Figure\,\ref{fig:3vel2} for three velocity ranges of 110--114, 114--118, and 118--128\,km\,s, where the LVC is shown in (b), while the HVC\,2 and intermediate velocity features are in (d) and (c), respectively (see the horizontal lines in Figure\,\ref{fig:lv}(b)).
The 8\,$\micron$ emission in this H{\sc ii} region consists of three distinct features.
The brightest feature is located at $(l,b)\sim(24\fdg40, 0\fdg04$--$0\fdg10)$ with an elongated shape along the north-south direction. 
We hereafter refer to this source ``H{\sc ii} region A1'', as labelled in Figure\,\ref{fig:3vel2}(a). 
H{\sc ii} region A1 is associated with a radio continuum source of size $\sim6$\,pc, and the 24\,$\micron$ emission is distributed to the west of the 8\,$\micron$ feature.
The second source, hereafter ``H{\sc ii} region A2'', is distributed to the west of H{\sc ii} region A1 by $\sim$3\,pc. 
It is smaller than H{\sc ii} region A1, with a size measured as $\sim3$\,pc, and is associated with the 24\,$\micron$ and 21\,cm emission.
The third 8\,$\micron$ feature (``H{\sc ii} region A3'') is located at $(l,b)\sim(24\fdg45, 0\fdg10)$ and has relatively weak emission.
H{\sc ii} regions A2 and A3 are cataloged as ``candidate'' H{\sc ii} regions G024.356+00.048 and G024.454+00.106, respectively, in the WISE Galactic H{\sc ii} region catalog \citep{and2014,mak2017}.
The WISE H{\sc ii} region catalog indicate the $f_{\rm \nu}$ measured with the MAGPIS 20\,cm and VGPS 21\,cm data as 3.1\,Jy and 4.3\,Jy, respectively, for H{\sc ii} regions A1, and 0.18\,Jy and 0.27\,Jy for H{\sc ii} regions A2, as summarized in Table\,\ref{tab:1} \citep{mak2017}.
The $N_{\rm ly}$ of H{\sc ii} regions A1 and A2 were then computed as $\sim10^{49.31}$--$10^{49.46}$\,s$^{-1}$ and $10^{48.08}$--$10^{48.25}$\,s$^{-1}$, respectively, in the same manner as adopted for N35 in Section\,3.2.1. 
The corresponding spectral types can be estimated as O4V--O5V and O8V--O8.5V, respectively.
As the 21\,cm emission is not detected in H{\sc ii} region A3, its $N_{\rm ly}$ cannot be measured.

\begin{figure}
 \begin{center}
  \includegraphics[width=13cm]{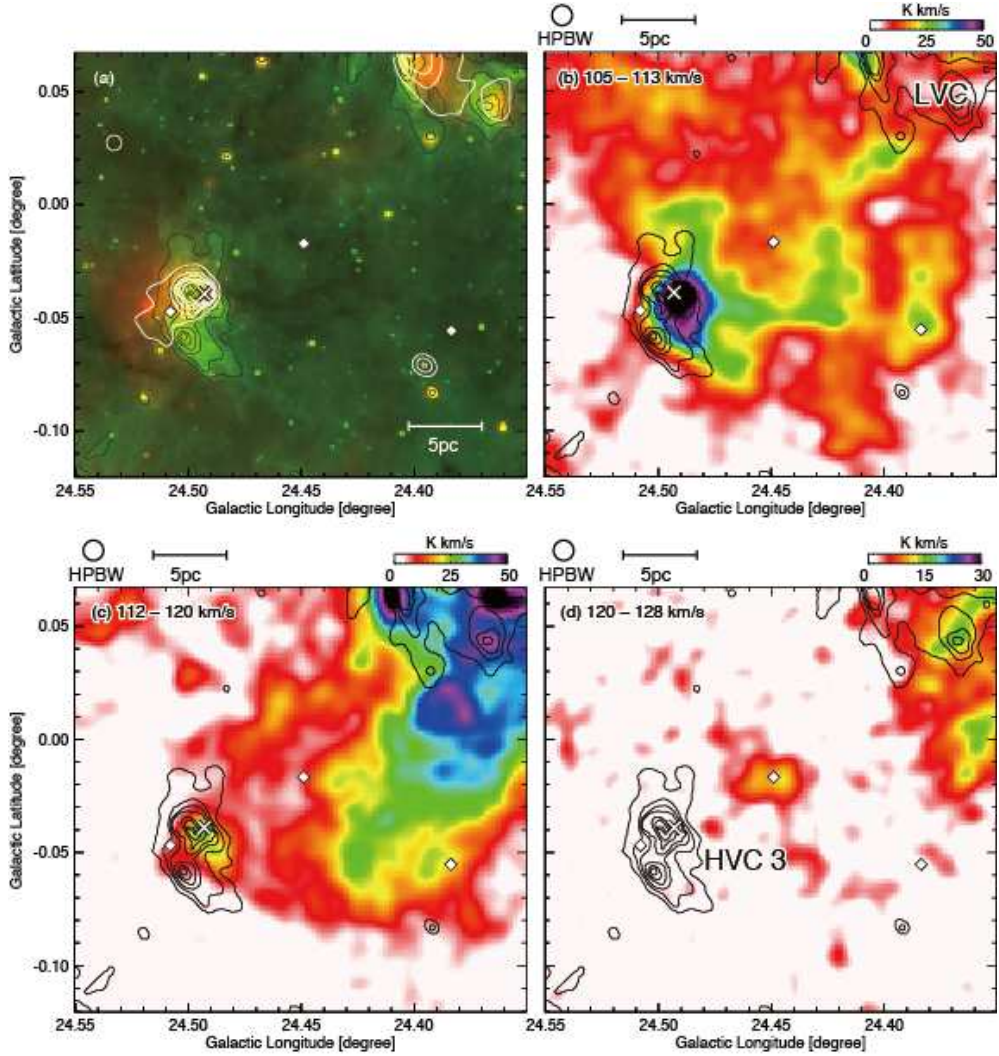}
 \end{center}
 \caption{Same as Figure\,\ref{fig:3vel1} but for H{\sc ii} region B. The velocity ranges in (b), (c), and (d) are 105--112\,km\,s$^{-1}$, 112--120\,km\,s$^{-1}$, and 120--128\,km\,s$^{-1}$, respectively.}\label{fig:3vel3}
\end{figure}

The 8\,$\micron$ emission of H{\sc ii} region A1 is traced by the bright CO emission in the LVC as seen in Figure\,\ref{fig:3vel2}(b), which is connected to a curved structure of the LVC elongated to the northwest.
A CH$_3$OH maser is detected at a local CO peak in this structure at $(l,b)\sim(24\fdg33, 0\fdg14)$.
The CO emission in the intermediate-velocity range in Figure\,\ref{fig:3vel2}(c) shows extended distribution centered on H{\sc ii} region A2 with a size of $\sim$8--9\,pc.
In this velocity range, another CO component is seen at $l\sim24\fdg43$ and $b>0\fdg10$, which corresponds to the southern half of the straight feature seen in Figure\,\ref{fig:3vel1}(b). 
H{\sc ii} region A3 is located to the east of the southern tip of this straight feature, implying interaction between these two.
Gas distribution of the HVC\,2 in Figure\,\ref{fig:3vel2}(d) shows elongated distribution toward the southwest, which has a CO peak at the corresponding direction of H{\sc ii} region A2.
Compared to the gas distribution in the LVC in Figure\,\ref{fig:3vel2}(b), the northern part of the HVC\,2 is surrounded by the curved feature of the LVC, indicating complementary distribution as shown in Figures\,\ref{fig:2vel}(c) and (f).
As seen in Figures\,\ref{fig:lv} and \ref{fig:mom1}, the LVC and HVC\,2 are connected with each other in velocity, and physical interaction between these two is suggested.

\subsubsection{H{\sc ii} region B}
CO distributions toward H{\sc ii} region B is shown in Figures\,\ref{fig:3vel3}(b), (c), and (d) for three velocity ranges, 105--113, 112--120, and 120--128\,km\,s$^{-1}$, respectively.
H{\sc ii} region B shown in Figure\,\ref{fig:3vel3}(a) has bright and compact 8\,$\micron$ peaks at $(l,b)\sim(24\fdg50, -0\fdg04)$ and $(24\fdg50, -0\fdg06)$. 
The 24\,$\micron$ and 21\,cm emission is enhanced toward the northern 8\,$\micron$ peak, and the 24\,$\micron$ is extended to the east by $\sim$3\,pc. 
We calculated the $N_{\rm ly}$ of H{\sc ii} region B as $10^{48.66}$--$10^{48.75}$\,s$^{-1}$ using the $f_{\rm \nu}$ of 0.68\,Jy at 20\,cm and 0.84\,Jy at 21\,cm from the WISE H{\sc ii} region catalog \citep{mak2017}, which correspond to a spectral type of O6.5V--O7.5V if we assume a single object (Table\,\ref{tab:1}).

As shown in Figure\,\ref{fig:3vel3}(b) in a velocity range of 105--113\,km\,s$^{-1}$, molecular gas in the LVC shows a ring-like morphology with a radius of $\sim6$\,pc.
The ring is closed in $^{12}$CO, whereas it is opened to the north in $^{13}$CO as shown in Figures\,\ref{fig:2vel}(c) and (f). 
The two 8\,$\micron$ sources of H{\sc ii} region B are located to the east and the southeast of a bright CO clump embedded within the eastern rim of the ring structure in the LVC. 
The CO clump is associated with a CH$_3$OH maser as depicted by cross in Figure\,\ref{fig:3vel3}, implying massive star formation.
Spatial distribution of the HVC\,3 is presented in Figure\,\ref{fig:3vel3}(d) for a velocity range of 120--128\,km\,s$^{-1}$. 
A {\it Spitzer} red source which is spatially coincident with the HVC\,3 suggest star formation in the HVC\,3.
The radius of the HVC\,3, $\sim$3\,pc, is consistent with the inner radius of the ring structure in the LVC, showing complementary distribution. 
The LVC and HVC\,3 are separated in velocity by the CO features with intermediate intensities shown in Figure\,\ref{fig:3vel3}(c).

\begin{table}
  \tbl{H{\sc ii} region properties }{
  \begin{tabular}{ccccccc}
  \hline
  \hline
    Region & H{\sc ii} region & $f_{\rm \nu}$ (20\,cm) & log $N_{\rm ly}$ (20\,cm) & $f_{\rm \nu}$ (21\,cm) & log $N_{\rm ly}$ (21\,cm) & Spectral type  \\
    && (Jy) & (log s$^{-1}$) & (Jy) & (log s$^{-1}$) & \\
    (1) & (2) & (3) & (4) & (5) & (6) & (7)\\
    \hline
    N35                       & G024.484+00.213 & 10.1 & 49.83 & 7.9 & 49.72 & $<$\,O3V \\
    H{\sc ii} region A1 & G024.392+00.072 &   3.1 & 49.31 & 4.3  & 49.46 & O4V--O5V\\
    H{\sc ii} region A2 & G024.356+00.048 &   0.18 &  48.08 & 0.27 & 48.25 & O8V--O8.5V \\
    H{\sc ii} region B   & G024.510-00.060  &   0.68 & 48.66 & 0.84 & 48.75  & O6.5V--O7V  \\
    \hline
  \end{tabular}}\label{tab:1}
  \begin{tabnote}
    (1) Name of the region in this paper. (2) Name of the H{\sc ii} region or H{\sc ii} region candidate. (3-4) The $f_{\rm \nu}$ and $N_{\rm ly}$ of the H{\sc ii} region measured using the MAGPIS 20\,cm data. The $f_{\rm \nu}$ was obtained from the WISE catalog web site \citep{mak2017}. (5-6) Same as (3-4) but for the VGPS 21\,cm data \citep{sti2006}. (7) Spectral type of the exciting source computed from $N_{\rm ly}$ listed in (3) and (5) assuming a single object \citep{mar2005}.
  \end{tabnote}
\end{table}

\section{Discussion}
In the previous sections we revealed that the Galactic infrared ring-like structure N35 is associated with a large GMC of a total molecular mass $\sim2.1\times10^6$\,$M_\odot$ at a distance of 8.8\,kpc. 
The GMC includes two other H{\sc ii} regions, H{\sc ii} region A and H{\sc ii} region B, which lay to the south of N35.
Our FUGIN CO $J$=1--0 data indicates that the GMC has two velocity components, the LVC and HVCs, and these two velocity components are connected with each other by the intermediate-velocity features.
In addition, the LVC and HVCs show spatially complementary distributions, although these are separated by 5--15\,km\,s$^{-1}$ in velocity.
These observational properties suggests physical interactions between the LVC and HVCs.

\subsection{The cloud-cloud collision model}

\begin{figure}
 \begin{center}
  \includegraphics[width=8cm]{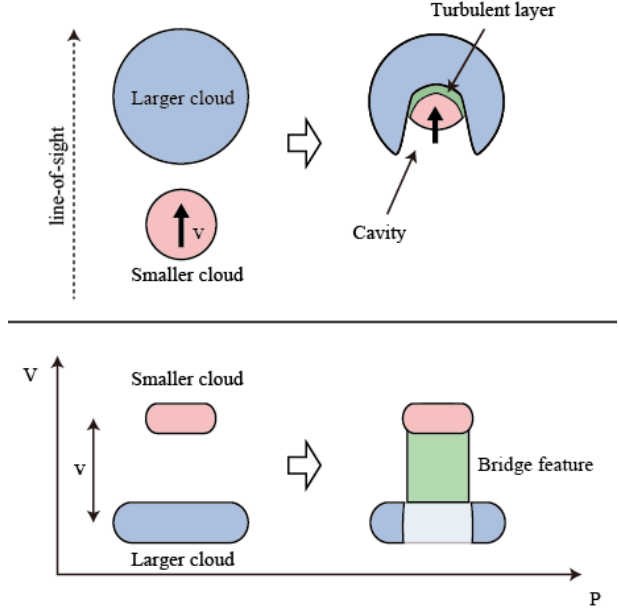}
 \end{center}
 \caption{Schematics of a collision between two clouds with different sizes before the collision (left) and during the collision (right), and a cartoon of position-velocity diagrams which show two colliding clouds and the bridge feature between the two clouds. Different color components in the collision schematics correspond to the different colors on the position-velocity diagrams.}\label{fig:bf1}
\end{figure}

We here propose a CCC model as an idea that explains the observational signatures in the N35 GMC. Recent studies on numerical calculations and molecular line observations revealed two observational signatures characteristic to CCC (e.g., \cite{tor2017ccc, fuk2017ccc1, haw2015b}). 
One is ``broad bridge feature in position-velocity diagrams'', and the other is ``complementary distribution between two clouds with different velocities''.
A broad bridge feature is gas at intermediate velocities between two clouds that are spatially coincident but separated in
velocity.
Figure\,\ref{fig:bf1} shows the schematics of CCC, as well as cartoons of the position-velocity diagram in which a broad bridge feature is included. When two clouds with different sizes collide, a smaller cloud drives into a larger cloud, resulting in a dense compressed layer, as the entirety of the smaller cloud undergoes collision quite quickly. The compressed smaller cloud continues to move into the larger cloud, forming a thin turbulent layer with velocities that are intermediate between the smaller cloud and the larger cloud. The intermediate-velocity gas is replenished as long as the collision continues. At a viewing angle parallel to the colliding axis, an observer sees two velocity peaks separated by lower intensity intermediate-velocity emission in the position-velocity diagrams, which corresponds to the broad bridge features. Several observational studies reported detections of broad bridge features in the CCC regions (e.g., \cite{tor2017ccc, tor2017, fuk2017ccc1, fuk2017ccc2}).

Another signature of CCC, complementary distribution, was discussed by \citet{tor2017} and \citet{fuk2017ccc1} based on comparisons between molecular line observations and numerical calculations of CCC \citep{tak2014, haw2015b, haw2015}.
When the smaller cloud drives into the larger cloud, it forms a cavity in the larger cloud, the size of which corresponds to that of the smaller cloud. If the observer has a viewing angle parallel to the collisional axis so that the two colliding clouds are spatially coincident along the line-of-sight, the larger cloud shows a ring distribution with an inner-radius that corresponds to the radius of the smaller cloud, forming a complementary distribution between these two clouds.

\begin{figure}
 \begin{center}
  \includegraphics[width=7cm]{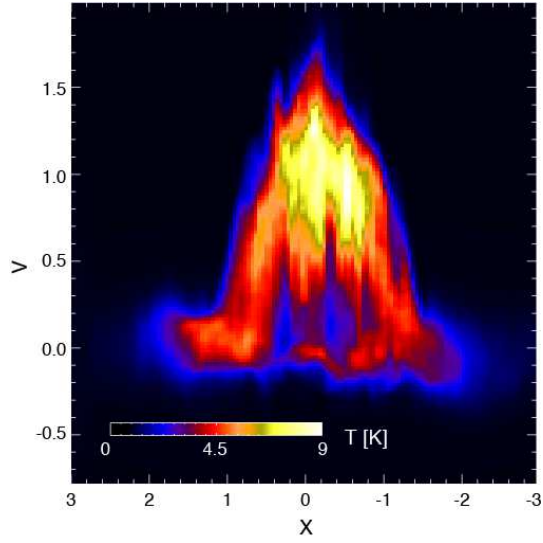}
 \end{center}
 \caption{The position-velocity diagram of the synthetic CO $J$=1--0 data produced by \citet{haw2015} based on the 10\,km\,s$^{-1}$ CCC model in \citet{tak2014}. The integration range in Y is set to be same with the size of the smaller cloud. The labels of the X and V axes are normalized with the radius of the small cloud as 3.5\,pc and the velocity difference between the small cloud and the large cloud as 4\,km\,s$^{-1}$, respectively. }\label{fig:bf2}
\end{figure}

As the cavity that forms in the larger cloud is seen as an intensity depression in the position-velocity diagram (see Figure\,\ref{fig:bf1}), \citet{tor2017ccc} discussed that two colliding clouds shows a ``V''-shaped gas distribution in the position-velocity diagram, and each side of the V-shaped structure is observed as a velocity gradient.
Figure\,\ref{fig:bf2} shows a position-velocity diagram of the synthetic CO $J$=1--0 cube data generated by \citet{haw2015} using the CCC model with a colliding velocity of 10\,km\,s$^{-1}$ calculated by \citet{tak2014}, where the viewing angle is set to parallel to the colliding axis. 
In Figure\,\ref{fig:bf2} the larger cloud and the smaller cloud can be seen around velocities of zero and one, respectively, and the CO emission overall shows an inverted V-shaped gas distribution.

\subsection{CCCs in the N35 GMC}
\begin{figure}
 \begin{center}
  \includegraphics[width=7cm]{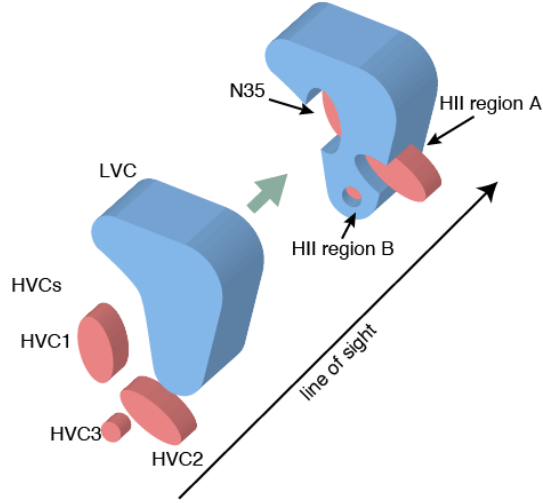}
 \end{center}
 \caption{Schematic picture of the cloud-cloud collisions in the GMC, where star formation is not included.}\label{fig:sch}
\end{figure}

We here discuss that the observed complementary distribution and intermediate velocity features between the 
LVC and HVCs can be interpreted by the CCC model.
Figure\,\ref{fig:sch} shows schematics of the collisions we propose here. 
We assume collisions between a large GMC extended for $\sim$30\,pc\,$\times$\,50\,pc and three smaller clouds at colliding
velocities of $\sim$10--15\,km\,s$^{-1}$, with an observer viewing angle almost parallel to the line-of-sight. 
The first case corresponds to the LVC, while the latter three are the HVCs (Figure\,\ref{fig:2vel}). 
We derived the masses of the HVCs\,1, 2, and 3 as about $1.3\times10^5$\,$M_\odot$, $0.3\times10^5$\,$M_\odot$, and $0.1\times10^4$\,$M_\odot$, respectively, from the $^{12}$CO integrated intensity map in Figure\,\ref{fig:2vel}(b), assuming an X-factor of $2\times10^{20}$\,cm$^{-2}$\,(K\,km\,s$^{-1}$)$^{-1}$ \citep{str1998}.

As presented in Figures\,\ref{fig:3vel1}--\ref{fig:3vel3}, the LVC has holes or rings near the individual H{\sc ii} regions. 
The western rim of N35 is surrounded by the CO emission in the LVC (Figure\,\ref{fig:3vel1}(b)), forming a hole, while H{\sc ii} region A is embedded within the western rim of the ring-like gas structure in the LVC, which is opened to the south (Figure\,\ref{fig:3vel2}(b)). 
H{\sc ii} region B is seen at the western rim of another ring-like structure in the LVC (Figure\,\ref{fig:3vel3}(b)).
As these holes/rings show complementary distribution with the HVCs, these can be interpreted as the cavities created by the collisions between the LVC and HVCs (see Figure\,\ref{fig:bf1}). 
\citet{fuk2017ccc1}  pointed out that two colliding clouds sometimes show displacement depending on on the inclination angle of the relative motion to the line-of-sight. 
However, the present complementary distributions seen in the three HVCs show no clear displacement. 
It is therefore suggested that the relative motion between the LVC and HVCs are almost parallel to the line-of-sight.

The velocity distribution of gas between the LVC and HVCs also support for the present CCC scenario. 
In N35 the velocity gradient of gas seen in the position-velocity diagram in Figure\,\ref{fig:lv}(a) can be interpreted as one side of the inverted V-shaped gas distribution in Figure\,\ref{fig:bf2}. 
This is the case for an offset collision, which means that the HVC\,1 is not perfectly coincident with the LVC along the line-of-sight.
The velocity gradients seen between the LVC and HVC\,2 in Figure\,\ref{fig:lv}(b) are similar to the inverted V-shaped gas distribution of the model data in Figure\,\ref{fig:bf2}. 
As the southwestern part of the HVC\,2 is not overlapping the LVC along the line-of-sight, it is likely that only the northern part of this cloud is colliding with the LVC, suggesting that the southern part of the HVC\,2 holds the initial condition of the cloud prior to the collision.
In H{\sc ii} region B, as the ring-like structure of the LVC is closed in the $^{12}$CO emission (Figures\,\ref{fig:mom1} and \ref{fig:2vel}), indicating that entirety of the HVC\,3 undergoes the collision.
As shown in the position-velocity diagram in Figure\,\ref{fig:lv}(c), the velocity distribution of the HVC\,3 connected to the LVC by the intermediate velocity feature resembles the cartoon position-velocity diagram in Figure\,\ref{fig:bf1}.

The timescale of the collision in each HVC can be calculated as the crossing time of the two colliding clouds. 
If we tentatively assume the size of the LVC and HVC\,1 along the line-of-sight as 30\,pc and 10\,pc, respectively, the timescale for the HVC\,1 to punch the LVC can be derived as (30\,pc + 10\,pc)\,/\,8\,km\,s$^{-1}$\,$\simeq$\,5\,Myr.
As the detection of the intermediate velocity emission indicates that the collision of the HVC\,1 still continues, the present timescale of the collision is probably shorter than this estimate. Similarly, the timescales of the collisions can be computed as less than (30\,pc + 10\,pc)\,/\,5\,km\,s$^{-1}$\,$\simeq$\,8\,Myr and (30\,pc + 3\,pc)\,/\,16\,km\,s$^{-1}$\,$\simeq$\,2.1\,Myr, respectively. 

It is unclear whether the HVCs could indeed penetrate the LVC in the future as results of the collisions, as the deceleration
of the collisions owing to momentum conservation,  as discussed by \citet{haw2015b}, is sometimes very effective.
However, the escape velocity $v_{\rm esc}$ from the N35 GMC can be calculated as $v_{\rm esc} = \sqrt{2 G M / r} \sim 26$\,km\,s$^{-1}$ by assuming a mass $M$ of $2.1\times10^6$\,$M_\odot$ and a radius $r$ of 25\,pc, which is much higher than the observed velocity separations between the LVC and HVCs, indicating gravitational binding between the LVC and HVCs. This implies that in the future the LVC and HVCs will be merged into one, becoming a single GMC, even if the HVCs will be able to penetrate the LVC.

\subsection{Ages of the H{\sc ii} regions in the N35 GMC}
It is of interest to investigate possibilities that the CCCs discussed in Figure\,\ref{fig:sch} triggered the high-mass star formation
in the N35 GMC. The formation timescale of an H{\sc ii} region is usually estimated using a D-type expansion model, and we here adopt the analytical model of the D-type expansion formulated by \citet{hos2006}
\begin{equation}
r_{\rm HII} (t) =  r_{\rm St} \left( 1 + \frac{7}{4} \sqrt{\frac{4}{3}} \frac{c_{\rm i} t}{r_{\rm St}} \right)^{4/7}, \label{eq:2} \\
\end{equation}
where $r_{\rm St}$ is the Str$\ddot{\rm o}$mgren radius calculated from $N_{\rm Ly}$ and ambient gas density $n_0$, and $c_{\rm i}$ is the isothermal sound speed, which corresponds to $\sim$11.5\,km\,s$^{-1}$ at  $T_{\rm e}$ of 8000\,K.

It is
noted that, based on the three-dimensional simulations of H{\sc ii} expansion within the clouds with fractal density distributions,
\citet{wal2012} discussed that the $r_{\rm HII}$ tracks the uniform density solution quite closely for all the test cases, suggesting that pc-scale mean gas density is a crucial parameter for the $r_{\rm HII}$ evolution. In the N35 GMC, the ridge feature seen at $l\sim24\fdg4$--$24\fdg5$ in Figures\,\ref{fig:wco}(a) and (b) have higher $N$(H$_2$) of (4--5)\,$\times10^{22}$\,cm$^{-2}$ as discussed in Section\,3.1. 
Given the width of the ridge feature of $\sim$6\,pc, the mean gas densities can be estimated as $\sim$2100--3200\,cm$^{-3}$. As N35 and H{\sc ii} region A are located close to the ridge feature, and the gas associated with H{\sc ii} region B has similar $N$(H$_2$) as the ridge feature, we here deemed the derived densities 2100--3200\,cm$^{-3}$ as $n_0$.

\begin{figure}
 \begin{center}
  \includegraphics[width=15cm]{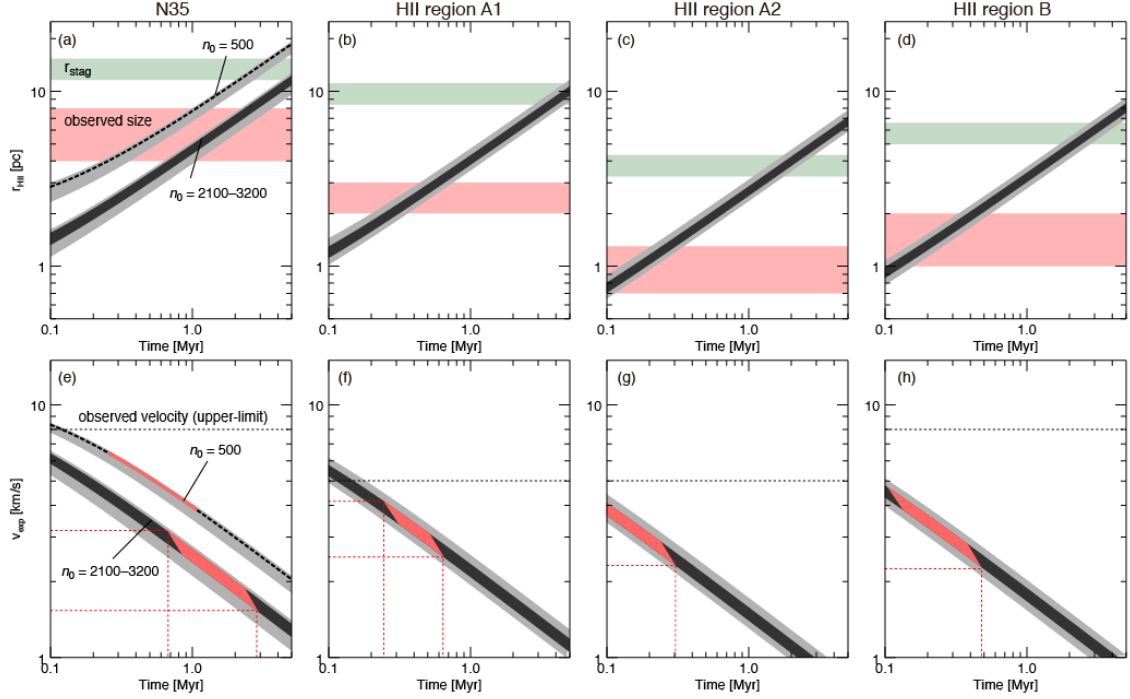}
 \end{center}
 \caption{Evolutionary tracks of the four H{\sc ii}  regions associated with the N35 GMC are presented using the D-type expansion equation provided by \citet{hos2006}, in which we assumed the $N_{\rm ly}$ listed in Table\,\ref{tab:1}. Black curves in the upper and lower panels show the $r_{\rm HII}$ and $v_{\rm exp}$ at $n_0=2100$--$3200$\,cm$^{-3}$ as functions of time in Myr, respectively, while the dashed black curve in panel (a) is the case for $n_0=500$\,cm$^{-3}$. The gray curves indicate the  $\pm50$\% uncertainties of the $r_{\rm HII}$ and $v_{\rm exp}$ curves. The radii of the H{\sc ii} regions measured from the THOR 21\,cm data are shown by filled red areas in the upper panels (Figure\,\ref{fig:rgb}, while the measured velocity dispersion toward the H{\sc ii} regions are indicated as horizontal dotted black lines in the lower panels. The red areas in the lower panels indicate the time and $v_{\rm exp}$ at which $r_{\rm HII}$ corresponds to the observed radii of the H{\sc ii} regions. The gray-green shades in the upper panels show the rstag of the H{\sc ii} regions (see the text).
 }\label{fig:dtype}
\end{figure}

In Figure\,\ref{fig:dtype} we show evolutionary tracks of the H{\sc ii} regions using the Hosokawa-Inutsuka model. 
The black curves in the upper and lower panels respectively indicate the full ranges of the radius of the H{\sc ii} region $r_{\rm HII}$ and the expanding velocity of the H{\sc ii} region $v_{\rm exp}$ as functions of time, which were computed with the $N_{\rm ly}$ obtained from the MAGPIS 20\,cm and VGPS 21\,cm data (Table\,\ref{tab:1}). 
We also plotted uniform $\pm50$\% uncertainties for the $r_{\rm HII}$ and $v_{\rm exp}$ as the gray curves in Figure\,\ref{fig:dtype}.
In each of the lower panels the filled area in red indicates the time and $r_{\rm exp}$ at which $r_{\rm HII}$  is consistent with the observed size of the H{\sc ii} region, which is shown
as the filled red areas in the corresponding upper panel.
As pointed out by \citet{bis2015}, the Hosokawa-Inutsuka model is an approximation of the model of \citet{rag2012b} at an early evolutionary stage of H{\sc ii} expansion. 
At a later time the Raga model, which takes into account the pressure acting from the neutral gas within the expanding shell on the ionized gas, indicates that the H{\sc ii} expansion is stagnated at $r_{\rm stag} = r_{\rm St} \left(\frac{8}{3}\right)^{2/3} \left(\frac{c_{\rm i}}{c_{\rm n}}\right)^{4/3}$, where $c_{\rm n}$ is the sound speed of the neutral gas, which was calculated assuming a neutral gas temperature of 30\,K from the molecular line observations of the Galactic H{\sc ii} regions (e.g., \cite{tor2011, fuk2016}).
The computed $r_{\rm stag}$ values are plotted as gray-green shades in Figure\,\ref{fig:dtype}(a)--(d), showing that the observed $r_{\rm HII}$ in all four H{\sc ii} regions are smaller than the computed $r_{\rm stag}$, and it is therefore suggested that the early-phase approximation of the Raga model, which corresponds to the Hosokawa-Inutsuka model, can be applied to these H{\sc ii} regions in the N35 GMC.

In Figure\,\ref{fig:dtype}(a) the curves of the Hosokawa-Inutsuka model indicate an evolutionary timescale of N35 as $\sim0.7$--3.0\,Myr, where we assumed the $r_{\rm HII}$ of N35 as 4--8\,pc by considering the uncertainty on the location of the exciting source in N35.
The corresponding $v_{\rm exp}$ measured as $\sim$1.5--3\,km\,s$^{-1}$ shown in Figure\,\ref{fig:dtype}(e) is also consistent with the observations, as the observed velocity dispersion toward N35 is as large as $\sim$8\,km\,s$^{-1}$ as indicated by the horizontal dotted black line in Figure\,\ref{fig:dtype}(e).
Figures\,\ref{fig:dtype}(b)--(d) show that H{\sc ii} region A1 has timescales of  0.2--0.6\,My, while H{\sc ii} region A2 and B indicate even shorter timescales; $<0.3$\,Myr and  $<0.5$\,Myr, respectively.
The estimated $v_{\rm exp}$ in each of these three H{\sc ii} regions is also consistent with the measured velocity dispersions (Figures\,\ref{fig:dtype}(f)--(h)).

The relatively longer timescale of N35 estimated in Figure\,\ref{fig:dtype} is due to the larger size of N35 and the assumption of the common $n_0$. 
As shown in Figure\,\ref{fig:2vel}, N35 is located at the eastern rim of the N35 GMC, and the eastern part of N35 appears to not be interacting with the N35 GMC.
In addition, \citet{bea2010} discussed that the infrared ring-like structures, including N35, have two-dimensional ring-like morphologies.
These observational signatures suggest that N35 has been evolving through the diffuse ISM surrounding the N35 GMC.
It is therefore preferable to assume lower $n_0$ in estimating the evolutionary timescale of N35, and if we tentatively assume $n_0$ of 500\,cm$^{-3}$, the estimated timescale of N35 is decreased to  $\sim$0.2--1.0\,Myr (see curves with dashed lines in Figures\,\ref{fig:dtype}(a) and (e)), which is rather similar to the figures in the other H{\sc ii} regions derived assuming $n_0$ of 2100--3200\,cm$^{-3}$.
As H{\sc ii} regions A1, A2, and B have smaller sizes than N35 and yet embedded within the N35 GMC, the assumption of $n_0$\,=\,2100--3200\,cm$^{-3}$ is rather reasonable.

\subsection{High-mass star formation triggered by CCCs}
In the previous subsection, the formation timescales of the H{\sc ii} regions in the N35 GMC were estimated as $\sim$1\,Myr or less, much shorter than the collisional timescales estimated in the three HVCs, suggesting that the exciting stars of the four H{\sc ii}regions were formed after the onset of the collisions. This is consistent with the hypothesis that the formation of these exciting stars were triggered by the CCCs between the LVC and HVCs. In the CCC model, dense gas clumps are formed within the compressed layer at the interface of the collision, at which the bridge features at intermediate velocities and/or complementary distribution between two clouds can be seen unless the cloud dispersal by the stellar feedback is significant. The dense clumps gains high mass accretion rates such as $10^{-4}$--$10^{-3}$\,$M_\odot$\,yr$^{-1}$ as demonstrated by \citet{ino2017}, which satisfies the theoretical requirement to overcome the radiation pressure feedback of the forming O star (e.g., \cite{wol1986, mck2003, hos2010}), and high-mass stars are finally formed at short timescales
of $\sim$0.1\,Myr.

The HVC\,1 in N35 appears to be overlapping the peak of the 21\,cm emission, at which the exciting source of N35 is possibly located, and it is consistent with the scenario that the high-mass star(s) in N35 was (were) formed at the colliding part between the LVC and HVC\,1. H{\sc ii} region A1 is located at the western rim of the ring-like structure of the LVC, while H{\sc ii} region A2 is seen at the center of the HVC\,2, suggesting that H{\sc ii} region A1 was formed on the side of the cavity created by the collision, while H{\sc ii} region A2 was born at the bottom of the cavity. Similar to H{\sc ii} region A1, it is suggested that H{\sc ii} region B, which is embedded in the east of the ring-like structure in the LVC (Figure\,\ref{fig:3vel3}), was formed at the eastern side of the cavity created in the LVC by the HVC\,3. Further observations with high spatial resolution at a 0.1-pc scale are necessary to reveal distributions and dynamics of the gas at the colliding regions between the LVC and HVCs, which will allow us to investigate the detailed physical process of the CCCs and related high-mass star formation.

In summary, the collisions between the LVC and HVCs have started since less than $\sim$0.1\,Myr ago, and the high-mass stars which energize the H{\sc ii} regions in the N35 GMC were formed at short timescales of $\sim$0.1\,Myr within the dense gas layer created through the strong compression at the colliding interface. As these three regions show the intermediate velocity features, i.e., the broad bridge features, in the position-velocity diagrams (Figure\,\ref{fig:lv}), indicating that the collisions still continue, further star formation including high-mass stars will possibly be triggered in the N35 GMC in the future. As shown in Figures\,\ref{fig:3vel1}--\ref{fig:3vel3}, associations of the CH$_3$OH masers as well as the {\it Spitzer} red sources with the N35 GMC around colliding parts lend more credence to the formation of stars in the colliding parts in the next generation.

\section{Summary}

The conclusions of the present study are summarized as follows.
\begin{enumerate}
\item We analyzed the CO $J$=1--0 data, which was obtained as a part of the FUGIN project using the Nobeyama 45-m telescope, in the Galactic infrared ring-like structure N35 and the two nearby H{\sc ii} regions, G024.392+00.072 (H{\sc ii} region A) and G024.510-00.060 (H{\sc ii} region B). 
We revealed that these three H{\sc ii} regions are associated with a GMC (the N35 GMC) which has a total molecular mass of $2.1\times10^6$\,$M_\odot$ at 8.8\,kpc.
\item Our CO data indicates that the N35 GMC has two velocity components, i.e., i.e., the LVC and HVC, between $\sim$110--126\,km\,s$^{-1}$. 
The majority of molecular gas in the N35 GMC is included in the LVC, having a large size of $\sim$30\,$\times$\,50\,pc at $\sim$110--114\,km\,s$^{-1}$, while the three HVCs (HVC\,1, HVC\,2, and HVC|,3) with smaller sizes were identified at $\sim$120--126\,km\,s$^{-1}$ around the three H{\sc ii} regions. 
The LVC has holes or rings around the H{\sc ii} regions, and these holes/rings are spatially coincident with the three HVCs, showing complementary distributions between the two clouds along the line-of-sight. 
In addition, the LVC and HVCs are connected in velocity by the CO emission with intermediate intensities around the interface of the complementary distribution. 
The intermediate-velocity features in N35 and H{\sc ii} region A show steep velocity gradients, while that in H{\sc ii} region B is coincident with the HVC\,3 along the line-of-sight.
\item We discussed that the observed complementary distributions and intermediate velocity features between the two velocity components can be interpreted by the collisions between the LVC and HVCs at velocities of $\sim$5--15\,km\,s$^{-1}$. Compared to the theoretical works on CCC, we assumed collisions between a GMC extended for $\sim$30\,pc\,$\times$\,50\,pc  and three smaller clouds. The former GMC corresponds to the LVC, while the latter are the HVCs. In this model, the gas at the intermediate velocities between the LVC and HVCs can be interpreted as a broad bridge feature, which is the gas in the thin
turbulent layer formed at the interface of the two colliding clouds, and the complementary distributions of gas indicate the cavities created in the LVC through the collisions.
\item The timescales of the collisions can be estimated as less than about 5\,Myr, 8\,Myr, and 2\,Myrin N35, H{\sc ii} region A, and H{\sc ii} region B, respectively. These figures are significantly larger than the ages of these H{\sc ii} regions, less than $\sim$1\,Myr, estimated assuming an analytical model of H{\sc ii} expansion, suggesting that the high-mass stars which energize these H{\sc ii} regions were formed by the triggering of the collisions.
\item Broad bridge features at intermediate velocities suggest that the collisions in the present GMC are still continuing. 
As the N35 GMC shows associations with CH$_3$OH masers and {\it Spitzer} red sources, further star formation triggered by the collisions are possibly occurring.
\end{enumerate}

\begin{ack}
The authors thank the anonymous referee for his/her helpful comments. 
This work was financially supported by Grants-in-Aid for Scientific Research (KAKENHI) of the Japanese society for the Promotion of Science (JSPS; grant numbers 15H05694, 15K17607, 24224005, 26247026, 25287035, and 23540277). 
Data analysis of the CO emissions was in part carried out on the open use data analysis computer system at the Astronomy Data Center, ADC, of the National Astronomical Observatory of Japan.
This work is based in part on observations made with the {\it Spitzer} Space Telescope, which is operated by the Jet Propulsion Laboratory, California Institute of Technology under a contract with NASA. This research is also based on observations with AKARI, a JAXA project with the participation of ESA. This research has made use of data obtained from the SuperCOSMOS Science Archive, prepared and hosted by the Wide Field Astronomy Unit, Institute for Astronomy, University of Edinburgh, which is funded by the UK Science and Technology Facilities Council. 
\end{ack}


\appendix

\section{Channel maps}
We present the velocity channel maps of the $^{12}$CO, $^{13}$CO, and C$^{18}$O emission of the N35 GMC in Figures\,\ref{fig:12channel}, \ref{fig:13channel}, and \ref{fig:18channel}, respectively, for a velocity range between 102 and 125\,km\,s$^{-1}$.
The contours indicate the 90\,cm radio continuum emissions.

\makeatletter
\renewcommand{\thefigure}{A\arabic{figure}}
\setcounter{figure}{0}
\makeatother

\begin{figure}
 \begin{center}
  \includegraphics[width=15cm]{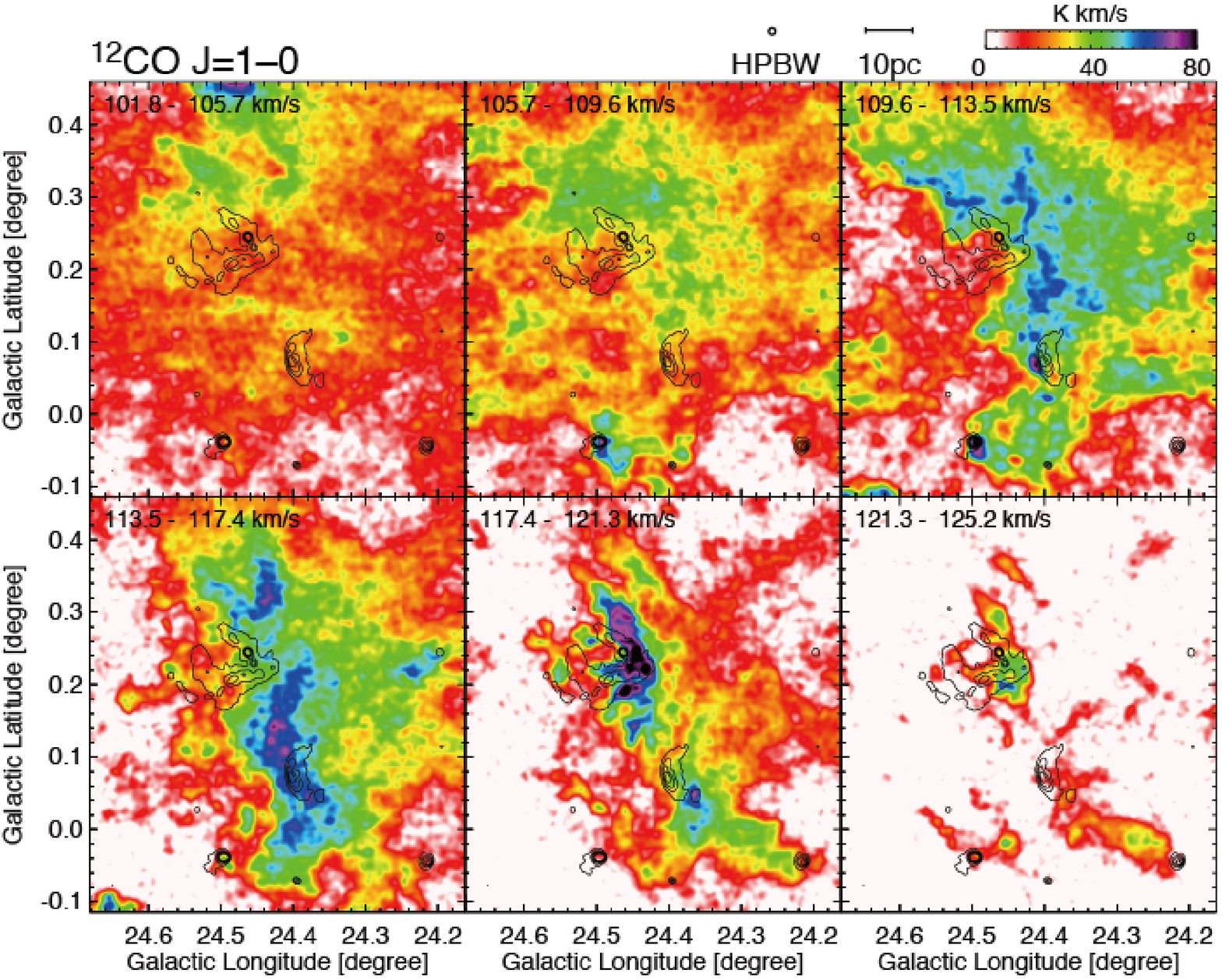}
 \end{center}
 \caption{Velocity channel maps of the $^{12}$CO emission. The contours indicate the 21\,cm radio continuum emission which start at 3\,$\sigma$ with steps of 1.5$\sigma$, which respectively corresponds to 45 and 22.5\,mJy\,beam$^{-1}$. }\label{fig:12channel}
\end{figure}

\begin{figure}
 \begin{center}
  \includegraphics[width=15cm]{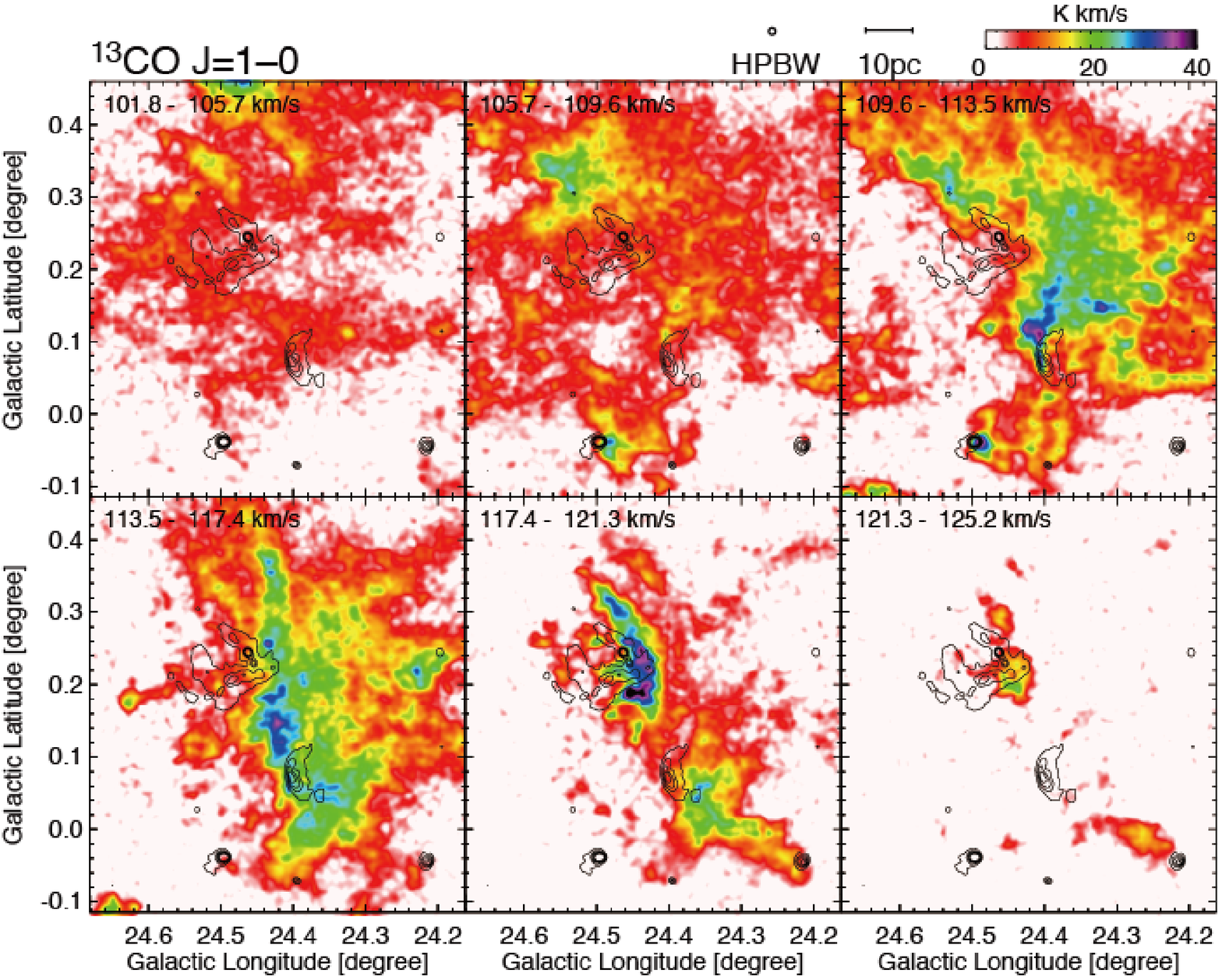}
 \end{center}
 \caption{Same as Figure\,\ref{fig:12channel} but for $^{13}$CO.}\label{fig:13channel}
\end{figure}

\begin{figure}
 \begin{center}
  \includegraphics[width=15cm]{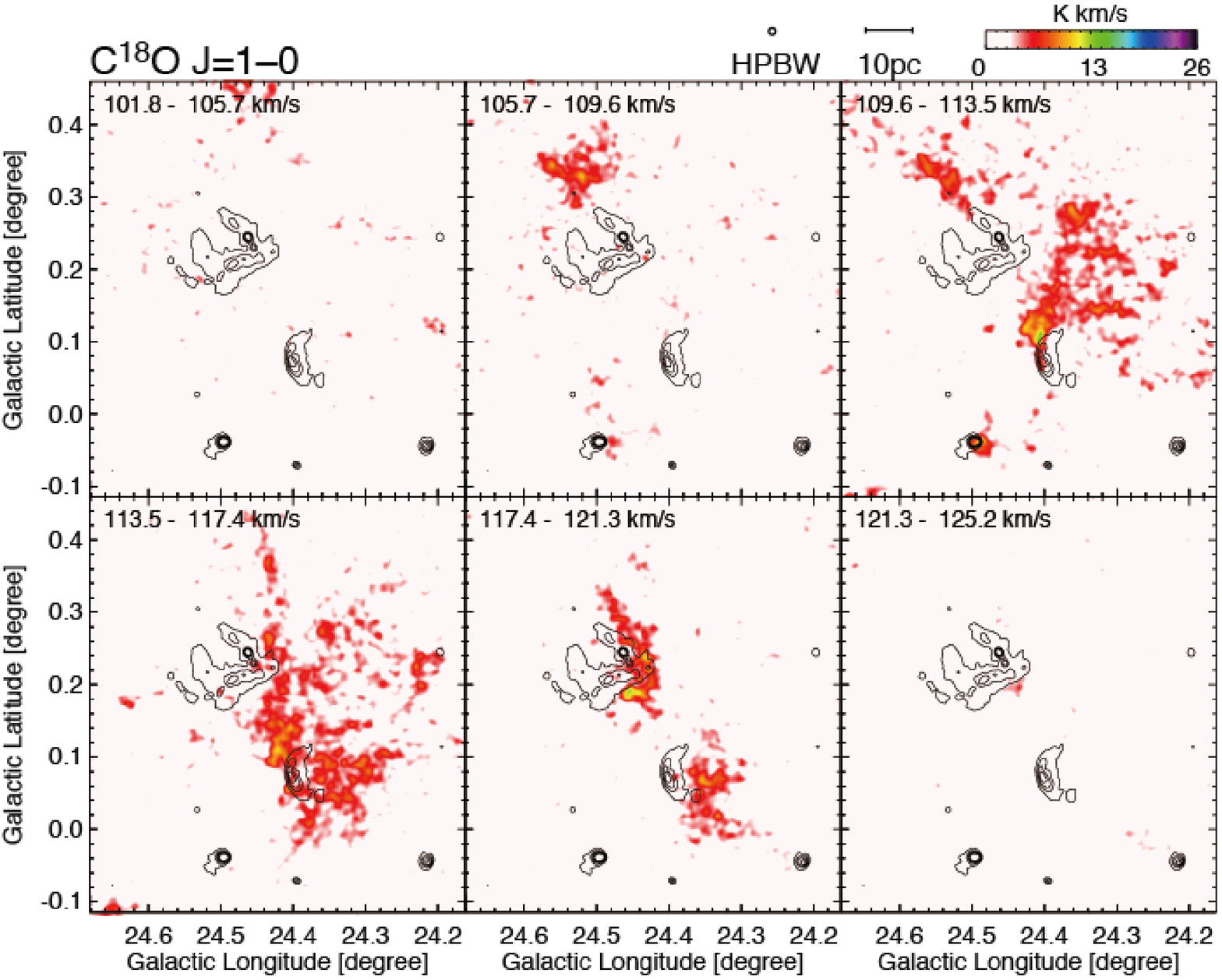}
 \end{center}
 \caption{Same as Figure\,\ref{fig:12channel} but for C$^{18}$O.}\label{fig:18channel}
\end{figure}

\end{document}